\documentclass[sigconf]{acmart} 
\AtBeginDocument{%
  }

\usepackage{booktabs}   
\usepackage{array}
\usepackage{tabularx}   
\usepackage{multirow}   
\usepackage{rotating}   
\usepackage{longtable}  
\usepackage{caption}    
\usepackage{enumitem}
\usepackage{dblfloatfix}
\usepackage[utf8]{inputenc}

\copyrightyear{2026}
\acmYear{2026}
\setcopyright{cc}
\setcctype{by}
\acmConference[CHI '26]{Proceedings of the 2026 CHI Conference on Human Factors in Computing Systems}{April 13--17, 2026}{Barcelona, Spain}
\acmBooktitle{Proceedings of the 2026 CHI Conference on Human Factors in Computing Systems (CHI '26), April 13--17, 2026, Barcelona, Spain}
\acmPrice{}

\begin{document}

\title{Facilitating Proactive and Reactive Guidance for Decision Making on the Web: A Design Probe with WebSeek}


\author{Yanwei Huang}
\affiliation{%
  \institution{The Hong Kong University of Science and Technology}
  \city{Hong Kong S.A.R.}
  \country{China}}
\email{yanwei.huang@connect.ust.hk}

\author{Arpit Narechania}
\affiliation{%
  \institution{The Hong Kong University of Science and Technology}
  \city{Hong Kong S.A.R.}
  \country{China}
}
\email{arpit@ust.hk}


\begin{abstract}
Web AI agents such as ChatGPT Agent and GenSpark are increasingly used for routine web-based tasks, yet they still rely on text-based input prompts, lack proactive detection of user intent, and offer no support for interactive data analysis and decision making. We present WebSeek, a mixed-initiative browser extension that enables users to discover and extract information from webpages to then flexibly build, transform, and refine tangible data artifacts–such as tables, lists, and visualizations–all within an interactive canvas. Within this environment, users can perform analysis–including data transformations such as joining tables or creating visualizations–while an in-built AI both proactively offers context-aware guidance and automation, and reactively responds to explicit user requests. An exploratory user study (N=15) with WebSeek as a probe reveals participants' diverse analysis strategies, underscoring their desire for transparency and control during human-AI collaboration.
\end{abstract}



\begin{teaserfigure}
  \centering
  \includegraphics[width=0.95\linewidth]{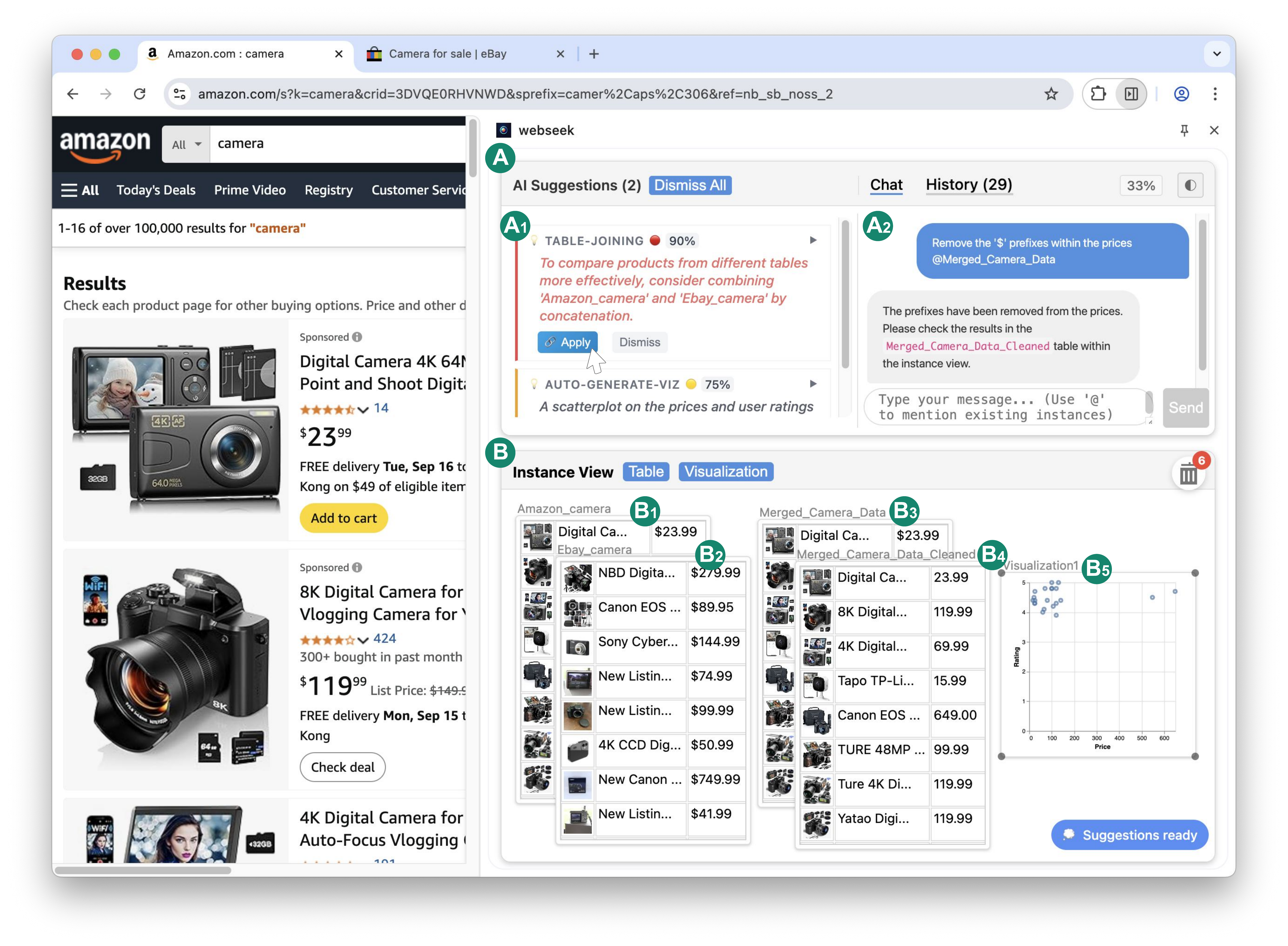}
  \vspace{-10mm}
  \caption{The WebSeek interface. (A) The AI Suggestions view, including a panel displaying proactive AI guidance (A1) and a Chat View (A2) for users to chat with the LLM to manage instances.
  (B) The instance view, a canvas holding the created data instances. They can either be data tables (B1-B4) or visualizations (B5).}
  \Description{The figure shows the WebSeek browser extension in the side panel. The view on the top left is the AI Suggestions view (A1), including a panel displaying several AI suggestions like "Table-joining" and "Auto-generate-viz". Next to it is the Chat view (A2), where users can send messages to the AI and the AI may respond in the same view. On the bottom of the system is the Instance View (B), which is a canvas containing multiple data instances like tables (B1: a table named "Amazon\_camera"; B2: a table named "Ebay\_camera"; B3: a table named "Merged\_Camera\_Data"; B4: a table named "Merged\_Camera\_Data\_Cleaned) and visualizations (B5: a scatterplot named "Visualization1"). }
  \label{fig:interface}
\end{teaserfigure}


\maketitle

\section{Introduction}

Modern web usage frequently involves complex, multi-step tasks that often culminate in a decision. Whether comparing consumer products, planning travel itineraries, booking movie tickets, or writing commercial reports, everyday web users\footnote{We refer to \emph{everyday web users} as individuals who predominantly use internet browsers to perform routine tasks and information-seeking activities without requiring specialized technical expertise.} often gather, synthesize, and analyze information from a variety of sources. 
These users often perform these tasks by juggling between multiple browser tabs and manually copy-pasting information into separate tools like spreadsheets or note-taking applications for subsequent analysis. 
Unfortunately, this workflow is often fragmented and hence laborious, imposing a high cognitive load on the user and creating friction in the overall process.

A key reason for this fragmented workflow is the lack of end-to-end data management tools. Data management is often characterized by the HCI community as an iterative, multi-step process (e.g., "Discover, Capture, Curate, Design, Create" by Muller et al.~\cite{muller2019data}), yet existing interactive web data extraction and wrangling tools~\cite{wrangler, rigel,xia2021ktabulator,dalvi2011automatic,mayer2015user,huang2023interactive} are predominantly built to support specific steps, leaving the overall workflow fragmented.
Recently, a new generation of web-based artificial intelligence (AI) agents (henceforth referred to as web agents), such as ChatGPT Agent~\cite{ChatGPT-Agent} and GenSpark~\cite{GenSpark}, have emerged, promising a more streamlined, end-to-end workflow, particularly for decision-making on the web. By leveraging the planning and reasoning abilities of large language models (LLMs), these agents can automate complex sequences of web actions based on natural language commands, such as browsing webpages, fetching data for analysis, and making web interactions on the user's behalf. In other words, these tools enable users to focus on decision making rather than data management, offering unprecedented efficiency for routine tasks on the web.

However, despite their advancements, these web agents fall short when it comes to supporting transparent and iterative data-driven analysis~\cite{chopra2025challenges, guo2024investigating}.
For instance, while these agents may produce final outputs or even expose fragments of their reasoning process, they often hide intermediate artifacts or present them in ways that are not easily usable by the user.
Even when outputs like tables are made visible, they are typically static, not interactive or customizable, hindering users from reusing or adapting the data analysis workflow\cite{chopra2025challenges}. 
We observe that this stems from a core design choice--a chatbot-like UI--which elevates text as the first-class citizen, not data. Without direct access to the underlying data and shared data representations, users are forced to articulate their requirements solely through text commands, an approach that can be inefficient, ambiguous, and particularly challenging for everyday web users~\cite{huang2023interactive} performing complex tasks. Essentially, there is a need to support these users in making informed and confident decisions on the web.

To address this need, we built WebSeek, an AI-powered mixed-initiative browser extension that treats data as a first-class citizen, enabling both the AI agent and the user to directly interact with it.
In particular, WebSeek shifts the traditional workflow towards building and refining tangible data artifacts, such as text snippets extracted from webpages and intermediate tables assembled from them.
Users can directly manipulate these artifacts (e.g., performing edits or data transformations) on an interactive canvas (\autoref{fig:interface} (B)), promoting efficient prototyping and enhancing transparency and flexibility. 
While this approach supports rich user interaction, achieving effective data-driven human–AI collaboration on the web presents new challenges: (1) the specific low-level data tasks needed to support complex, open-ended decision making on the web are ill-defined; and (2) determining when, what, and how to provide AI guidance to optimize the trade-off between efficiency and control remains a non-trivial design problem.
To address these challenges, we curate a principled design space of mixed-initiative assistance for data tasks on the web that formalizes AI intervention to support the user's goals. Guided by this design space, WebSeek enables direct manipulation augmented by a mixed-initiative AI. This AI both reactively executes user commands on demand and proactively suggests context-aware actions (\autoref{fig:interface} (A)), ranging from batch data extraction to complex table joins.

Using WebSeek as a design probe, we conducted a user study (N=15) involving two representative data-driven decision making scenarios: (1) fact-checking a news story and (2) product research. We found that WebSeek helped participants finish complex decision making tasks successfully while leading to high user confidence and sense of agency and control. 
Participants exhibited a diverse set of workflows with varying degree of reliance on AI guidance and direct manipulation. Overall, they found direct manipulation indispensable and perceived proactive and reactive guidance in WebSeek as reliable and helpful.
Drawing on these observations, we discuss implications for the design of future mixed-initiative data analysis systems.
Our contributions include:

\begin{itemize}

    \item A mixed-initiative browser extension, WebSeek, tailored for data-driven decision making.

    \item A framework for providing proactive and reactive AI guidance during web-based data-driven decision making, covering the timing, scope, and presentation of guidance. 
    
    \item Findings from an exploratory user study using WebSeek as a probe to understand how users perform two everyday tasks on the web.
    
    \end{itemize}
\section{Related Work}
\subsection{AI-powered Web Agents}
The recent proliferation of Large Language Models (LLMs) and AI agents has catalyzed a new paradigm of web automation. 
Moving beyond traditional programmatic scripting, numerous modern web agents have demonstrated impressive capabilities in streamlining web workflows using natural language.
From a user-interface perspective, these agents generally fall into three classes: (1) chatbot-style interfaces--in which an LLM can access the web and external tools to fulfill user requests (e.g., ChatGPT Agent~\cite{ChatGPT-Agent}, GenSpark~\cite{GenSpark},  AnythingLLM~\cite{AnythingLLM}); (2) browser extensions or forked browsers--that fetch page context and provide in-situ suggestions and actions, navigating users to quick chats with LLMs in a chatroom (e.g., Harpa.AI~\cite{Harpa-ai}, GenSpark Browser~\cite{GenSpark-Browser}, Merlin.AI~\cite{Merlin-ai}); and (3) visual agents--that execute end-to-end web automation workflows in a virtual machine or local environment while exposing the process visually for inspection and control (e.g., Skyvern~\cite{Skyvern}, Manus.AI~\cite{Manus-ai}, AgenticSeek~\cite{AgenticSeek}). 
These systems operate with a high degree of goal-driven autonomy, leveraging a suite of tools for web search, navigation, processing, and interaction to plan and execute complex tasks~\cite{he2024webvoyager, agentic-web, ma2023laser, webagent-survey, abuelsaad2024agent, wang2025mobile}. Some of these systems even decompose workflows into specialized subagents or incorporate multimodal capabilities for richer data collection and analysis~\cite{song2024mmac, zhang2023you}.

While this paradigm is powerful, its autonomous nature presents limitations for data-driven decision making, particularly for high-stakes decisions where users desire agency~\cite{shao2025future, 2019-agency-automation}. For instance, an end-to-end workflow often hides the intermediate data artifacts and manipulation steps, risking producing outcomes that are ungrounded, biased, or not personalized to the user's nuanced goals~\cite{cronin2024understanding}. 
Although tools such as AgenticSeek~\cite{AgenticSeek} and Skyvern~\cite{Skyvern} visualize the agent's decision making process, the data artifacts are often presented as static snapshots, limiting users' ability to interact with and manipulate them directly.
Furthermore, while natural language is a powerful input modality for initiating tasks, it can become a bottleneck during the analysis process. When users need to customize or modify a data transformation pipeline, they must revert to articulating these complex needs solely through text. As prior work has shown, text is often not the ideal modality for specifying the nuances of data transformation, which can be more directly and unambiguously expressed through direct manipulation~\cite{srinivasan2021collecting, huang2023interactive, masson2024directgpt,shen2025prompting}.
In contrast, WebSeek proposes a data-driven and mixed-initiative paradigm, allowing users to work around tangible data artifacts for better transparency and human agency.

\subsection{Web Data Extraction and Analysis}
Extracting structured data from semi-structured webpages has been a long-standing challenge in HCI and data management research~\cite{doan2009information}. Early approaches relied on programmatic scraping libraries like Beautiful Soup~\cite{beautiful-soup} or custom wrappers~\cite{dalvi2011automatic, ortona2015wadar}, which are powerful but require significant programming expertise. To address this, a number of interactive tools have been proposed, leveraging paradigms such as Programming by Example (PBE) and Programming by Demonstration (PBD), enabling non-programmers to create data extractors through making data examples~\cite{le2014flashextract, inala2017webrelate} or demonstration the procedure~\cite{barman2016ringer,chasins2018rousillon, pu2022semanticon}. 
Given the extracted data, web data analysis, however, often happens in a different environment. A number of commercial tools and research prototypes have been developed for data profiling~\cite{profiler}, wrangling and cleaning~\cite{wrangler,trifacta,openrefine}, and visualization~\cite{satyanarayan2014lyra, zong2020lyra, satyanarayan2016vega}.


Despite technical advances, web data extraction and analysis still face two key limitations. First, data extraction and analysis are often conducted as separate processes, lacking a unified system or interface that allows users to seamlessly transition between these stages and engage in iterative exploration. Second, providing real-time, proactive guidance throughout the entire data analysis process remains challenging. Prior work on guidance for data wrangling~\cite{wrangler,chopra2023cowrangler,zhou2025xavier} and visualization~\cite{wongsuphasawat2015voyager,wongsuphasawat2017voyager} mostly bases recommendations on data characteristics (e.g., quality issues), though some visualization tools infer user intent from tasks and interactions to suggest charts~\cite{cao2022visguide,epperson2022leveraging}. 
However, most prior systems infer the user's intent from their current action (e.g., a single selection or edit) within an individual stage of the overall pipeline (e.g., extraction, wrangling, or cleaning), with limited exploration of using longer interaction histories to provide proactive guidance across stages.
WebSeek addresses the first challenge by introducing a data-centric human--AI collaboration paradigm within a browser extension that unifies these stages. It further investigates when, what, and how to provide proactive AI guidance grounded in user interactions to tackle the second challenge.

\subsection{Proactive Guidance for Data-Driven Decision Making}

Providing guidance as users interact with a system is widely recognized as a key approach to lowering learning barriers, improving task outcomes, and optimizing process efficiency~\cite{narechania2024dissertation, zhao2025proactiveva}. Building on this idea, a recent trend emphasizes \emph{proactive} guidance: instead of requiring users to request help, the system anticipates needs and offers timely suggestions to streamline workflows, reduce cognitive load, and strengthen user trust~\cite{peng2019design,proactive-trust,guo2011proactive, paden2023biasbuzz}. These methods typically infer user intent from the interaction history and data context (e.g., quality, provenance, and usage information), and then produce recommendations via rule-based or learned models~\cite{narechania2023datapilot, narechania2023datacockpit, cui2019datasite, learning-vis-recommendation}. With the rise of powerful LLMs, proactive AI agents that provide real-time guidance have been adopted across various domains such as programming~\cite{Copilot,zhou2025xavier,need-help}, data analytics~\cite{zhao2025proactiveva}, and autonomous driving~\cite{meck2024may}.

Despite this promise, improperly designed proactive guidance can be detrimental. Research has shown that ill-timed or irrelevant suggestions can disrupt a user's workflow and lead to negative experiences~\cite{pu2025assistance,what-can-you-do}. Moreover, users who desire agency and control may perceive proactive suggestions as intrusive, choosing to ignore them or use them only as a ``last resort''~\cite{guo2011proactive}. This highlights a critical design tension: proactive guidance must carefully balance its potential utility against the cognitive cost of interruption. Addressing this tension requires a principled approach to three core questions: \textit{what} content to suggest, \textit{when} to trigger it, and \textit{where} in the interface to display it. While researchers have proposed general design guidelines or studied a dimension of these questions ~\cite{narechania2025provenancelens, viswanath2025insights, pu2025promemassist, narechania2025agentic,guidance-typology}, there is a lack of frameworks that systematically address all these questions in specific domains. 
WebSeek contributes to this space by proposing and instantiating a detailed design space for proactive assistance in the context of data-driven sensemaking on the web, providing a concrete model for creating effective, non-disruptive web data analysis tools.

\section{A framework for mixed-initiative assistance in data tasks on the web}

WebSeek envisions enhancing data-driven decision making on the web with efficient human-AI collaboration centered on tangible data instances. However, during the design process, we soon observed two research questions that are crucial to the design decisions. First, \textbf{what kinds of data tasks are involved in the process of data-driven decision making on the web?} Despite the rich literature on general data analysis or decision making process~\cite{muller2019data, kandogan2014data,kandel2012enterprise,Futz-mosey}, existing models need to be properly adapted to accommodate the nuanced web environment, accounting for the specific, low-level actions users may perform.
Second, having identified these specific tasks, a natural follow-up question emerged: \textbf{how can an AI proactively and reactively assist users across this diverse set of tasks in a helpful and non-intrusive manner?} This question exposes the central tension in designing any mixed-initiative system: balancing the potential benefits of AI automation against the cognitive cost of interruption and the risk of undermining user agency~\cite{pu2025assistance, 2019-agency-automation}. The optimal way for an AI to assist with a small, well-defined task like autocompleting a column is fundamentally different from how it should suggest a complex, strategic action like joining two tables or generating a new visualization. A one-size-fits-all approach to assistance would inevitably lead to a frustrating user experience.

To address these questions, we envisioned a structured design space that would serve as the conceptual backbone for WebSeek's AI capabilities. To develop it, the authors of this paper, who are also experienced data analysts themselves, initiated a draft of the space and revised it through multiple rounds of discussion. During the process, we particularly focused on three core research questions summarized by us: a) When should the AI guidance be triggered?  b) What should the AI guidance include? c) How should the AI guidance be communicated to the users? In addition, we reflected on the space by reviewing well-known theories and principles on mixed-initiative and general UI design~\cite{horvitz1999principles, mackenzie2018fitts} after the discussion.

\subsection{Proactive Assistance} \label{proactiveassistance}

The framework, summarized in \autoref{tab:space}, provides a principled approach to designing proactive AI assistance for web data tasks. It answers the first question by adapting the empirically-grounded taxonomy of enterprise data analysis from Kandel et al.~\cite{kandel2012enterprise} to the specific context of the web, categorizing tasks into four stages: Discovery, Data Extraction \& Wrangling, Data Profiling \& Cleaning, and Data Modeling \& Visualization\footnote{The words ``Extraction'', ``Cleaning'', and ``Visualization'' do not exist in the original space by Kandel et al. but are mentioned in the subcategories. These terms are commonly used in data science and we add them to the category names to better align with web tasks. }. We have also summarized a list of features by adapating the subtasks in their taxonomy to the context of web data analysis. Crucially, the framework answers the second question by defining the scope and modality of AI assistance for each task. We distinguish between \textbf{micro} suggestions, which are small, efficiency-focused actions that accelerate a user's current activity (e.g., autocompleting data), and \textbf{macro} suggestions, which are larger, strategic actions that propose a new direction (e.g., creating a chart). Our guiding principle is to match the suggestion's scope to its presentation modality: micro suggestions are presented \textbf{in-situ}, directly within the user's focus, while macro suggestions are presented \textbf{peripherally} (i.e., ex-situ) in a dedicated panel to avoid disrupting the user's flow. Finally, we heuristically decided the user triggering signal, the potential proactive actions of AI, and the scope and modality of the guidance, with the validity examined iteratively through our design process.


\subsection{Reactive Assistance and User Manipulation}

While proactive assistance offers efficiency and automation, it is equally important to allow users to seek AI assitance reactively or manually interact in the system to guarantee user agency~\cite{2019-agency-automation}. Similar to the proactive assistance, the suitability of an interaction also varies between tasks. For instance, selecting specific elements on a webpage is naturally suited to direct manipulation, whereas extracting a list of data from webpages might be more efficiently expressed through demonstration or examples. Meanwhile, it is crucial for us to figure out how interactions can be designed around the tangible data artifacts that we envision our system to focus on. These factors motivate us to construct a design space that maps tasks and features to feasible interaction modalities.

\newlength{\catcol} \setlength{\catcol}{2cm}
\newlength{\featcol} \setlength{\featcol}{2.5cm}
\newlength{\trigcol} \setlength{\trigcol}{4.8cm}
\newlength{\actcol} \setlength{\actcol}{4.8cm}
\newlength{\scocol} \setlength{\scocol}{2.5cm}

\begin{table*}[!t]
\centering
\caption{A Framework for Proactive AI Assistance in The Workflow of Data-Driven Decision Making on the Web}
\label{tab:space}
\begin{tabular}{@{} m{\catcol} p{\featcol} p{\trigcol} p{\actcol} p{\scocol} @{}}
\toprule
\textbf{Task Category} & \textbf{Feature} & \textbf{User Trigger} & \textbf{Proactive AI Action} & \textbf{Scope \& Modality} \\
\midrule

\multirow{1}{=}{\raggedright 1. Discovery}
& Webpage Suggestion
& User creates a new workspace with a title/text indicating a research goal (e.g., \textit{"Best Phones Under \$800"}).
& Offers a list of relevant review sites, e-commerce pages, or articles to start the investigation.
& Macro / Peripheral \\
\midrule

\multirow{7}{=}{\raggedright 2. Data Extraction \& Wrangling}
& Element selection
& User selects a second element that is a positional or structural sibling to the first.
& Highlights all similar elements on the page and shows an overlay: \textit{"Select all (15) matching items?"}
& Micro / In-Situ or Macro / Peripheral \\
\cmidrule(l){2-5}

& Schema inference
& User extracts data into a new table but leaves column headers blank or default and clicks away.
& Analyzes content and places suggested headers as ghost text in the header cells.
& Micro / In-Situ \\
\cmidrule(l){2-5}

& Batch extraction
& User drags a second, structurally similar item from the same list/table onto the system.
& Detects the pattern, highlights remaining items, and prompts: \textit{"Extract all (20) rows?"}
& Micro / In-Situ \\
\cmidrule(l){2-5}

& Row/Col Autocomplete
& User establishes a pattern by manually filling at least two cells in a new column (e.g., combining first/last name).
& Infers the pattern and populates the rest of the column with ghost text, which can be accepted with Tab.
& Micro / In-Situ \\
\cmidrule(l){2-5}

& Computed columns
& User creates a table with columns having a clear mathematical relationship (e.g., 'Price', 'Quantity'), or user performs calculations in a cell of a new column.
& Suggests a new column with the computed formula in the Suggestions Panel.
& Micro / In-situ or Macro / Peripheral \\
\cmidrule(l){2-5}

& Sorting \& Filtering
& User manually applies the same filter/sort to two different but structurally similar tables.
& Suggests creating a similar rule that can be applied to new tables with one click.
& Macro / Peripheral \\
\cmidrule(l){2-5}

& Joining tables
& User has two or more tables on the canvas with a matching column (e.g., 'Product\_ID').
& Suggests joining the tables via the Suggestions Panel: \textit{"Join Table A and B on 'Product\_ID'?"}
& Macro / Peripheral \\
\midrule

\multirow{4}{=}{\raggedright 3. Data Profiling \& Cleaning}
& Entity Resolution
& User manually edits two cells in a column to be consistent (e.g., "USD" to "\$").
& Scans the column, identifies variations, and suggests a batch normalization.
& Micro / In-Situ \\
\cmidrule(l){2-5}

& Remove extraneous
& User manually removes the same character(s) from two cells (e.g., deleting "Sponsored" text).
& Offers to remove the same characters from all other cells where the pattern is found.
& Micro / In-Situ \\
\cmidrule(l){2-5}

& Fill missing values
& User has a table with several empty cells in a numerical column, or user fills in an empty cell.
& Suggests strategies in the Suggestions Panel: \textit{"Fill 8 missing values using the column average?"}
& Micro / In-Situ or Macro / Peripheral \\
\cmidrule(l){2-5}

& Data type correction
& A column contains mostly numbers, but some cells are text (e.g., "N/A"). User deletes one text value.
& Identifies non-numeric values and suggests a batch action, e.g., \textit{"Replace all 5 text values with '0'?"}
& Micro / In-Situ \\
\midrule

\multirow{3}{=}{\raggedright 4. Data Modeling \& Visualization}
& Auto-visualizations
& User selects a table artifact with at least one categorical and one numerical column.
& Suggests a relevant visualization in the Suggestions Panel, e.g., \textit{"Create a Bar Chart?"}
& Macro / Peripheral \\
\cmidrule(l){2-5}

& Suggest alternative charts
& User creates a visualization that is poorly suited for the data, or user modifies the chart specification.
& Suggests a more suitable alternative in the Suggestions Panel.
& Macro / Peripheral \\
\cmidrule(l){2-5}

& Interactive filtering
& User manually selects a subset of data in a table linked to a visualization.
& Suggests turning the one-off selection into a permanent, interactive filter control.
& Macro / Peripheral \\

\bottomrule
\end{tabular}
\end{table*}

\begin{table*}[t]
    \centering
    \caption{Mapping of Features to Interaction Modalities. "O" indicates existing literature applies the corresponding interaction to the task. "X" indicates that no existing literature is identified or there is no need for applying the interaction on the feature. Horizontal lines indicate the boundaries of merged cells. Cells of "O" in the \textit{Textual} and \textit{Manip.} columns  represent general interactions like chatting with LLMs using textual prompts and manipulating data in software like browser and spreadsheet tools respectively, hence no citations are provided.  (Explanation for the abbreviations: Demo - Programming by demonstration; Example: Programming by example; DSL - Domain-specific language; Textual - Prompting LLMs with textual prompts; Visual - Prompting LLMs with visual prompts; Manip. - Direct manipulation.)}
    \label{tab:interactions}
    \renewcommand{\arraystretch}{1.2} 

    \begin{tabularx}{\textwidth}{@{} >{\centering}p{2cm} >{\centering\arraybackslash}X c c c c c c c @{}}
        \toprule
        & & \multicolumn{7}{c}{\textbf{Interactions}} \\
        \cmidrule(lr){3-9}
        \textbf{Task Category} & \textbf{Feature} & \textbf{Coding} & \textbf{Demo.} & \textbf{Example} & \textbf{DSL} & \textbf{Textual} & \textbf{Visual} & \textbf{Manip.} \\
        \midrule
        
        \multirow{1}{2cm}{\centering\textbf{Discovery}} & Suggest  webpages & X & X & X & X & O & X & O \\
        \midrule

        \multirow{7}{2cm}{\centering\textbf{Data Extraction \& Wrangling}}
        & Element selection & \multirow{2}{1cm}{\centering X} & \multirow{2}{1cm}{\centering X} & \multirow{2}{1cm}{\centering X} & \multirow{2}{1cm}{\centering X} & X & \multirow{3}{1cm}{\centering X} & \multirow{7}{1cm}{\centering O} \\
        \cmidrule(lr){7-7}
        & Schema inference & & & & & \multirow{6}{1cm}{\centering O} & & \\
        \cmidrule(lr){3-6}
        & Batch extraction & O\cite{laender2002brief} & O\cite{barman2016ringer} & O\cite{laender2002debye} & O\cite{arifanto2018domain} & & & \\
        \cmidrule(lr){3-6} \cmidrule(lr){8-8}
        & Autocomplete & \multirow{4}{1cm}{\centering O\cite{pandas}} & \multirow{4}{1cm}{\centering O\cite{wrangler}} & \multirow{4}{1cm}{\centering O\cite{foofah}} & \multirow{4}{1cm}{\centering O\cite{rigel}} & & \multirow{4}{1cm}{\centering O\cite{hesenius2025draw}} & \\
        & Computed columns & & & & & & & \\
        & Sorting and Filtering & & & & & & & \\
        & Joining tables & & & & & & & \\
        \midrule

        \multirow{4}{2cm}{\centering\textbf{Data Profiling \& Cleaning}}
        & Entity Resolution & \multirow{4}{1cm}{\centering O\cite{pandas}} &  \multirow{4}{1cm}{\centering O\cite{wrangler}} & \multirow{4}{1cm}{\centering O\cite{foofah}} & \multirow{4}{1cm}{\centering O\cite{rigel}} & \multirow{4}{1cm}{\centering O} & \multirow{4}{1cm}{\centering O\cite{hesenius2025draw}} & \multirow{4}{1cm}{\centering O} \\
        & Remove characters & & & & & & & \\
        & Data type correction & & & & & & & \\
        & Fill missing values & & & & & & & \\
        \midrule

        \multirow{3}{2cm}{\centering\textbf{Data Modeling \& Visualization}}
        & Auto-generate viz. & \multirow{3}{1cm}{\centering O\cite{D3-js}} & \multirow{3}{1cm}{\centering O\cite{saket2016visualization}} & O\cite{shi2022supporting} & \multirow{3}{1cm}{\centering O\cite{satyanarayan2016vega}} & \multirow{3}{1cm}{\centering O} & \multirow{3}{1cm}{\centering O\cite{teng2021sketch2vis}} & \multirow{3}{1cm}{\centering O} \\
        \cmidrule(lr){5-5}
        & Suggest alt. charts & & & \multirow{2}{1cm}{\centering X} & & & & \\
        & Int. filt./highlighting & & & & & & & \\

        \bottomrule
    \end{tabularx}
\end{table*}

To systematically understand this design landscape, we surveyed prior work to identify the spectrum of interaction modalities that have been applied to web data tasks, ranging from imperative approaches like Coding and domain-specific languages (DSL) to more intuitive methods like programming by Demonstration~\cite{cypher1993watch} or Example~\cite{halbert1984programming}, and modern approaches like textual prompt-based (natural language) \cite{androutsopoulos2005natural} and visual prompt-based (annotation and sketches)~\cite{yang2023fine} interactions.

Table \ref{tab:interactions} presents the result space, mapping the data features from our design space against these seven distinct interaction modalities. Note that textual prompts and direct manipulation offer the broadest and most intuitive coverage for various web-based tasks. This informed WebSeek's data instance-oriented design: the data instances serve as a universal interaction unit which, ideally, can be directly manipulated to demonstrate user intent or build data examples, while being manageable through other interaction modalities. 

\section{Design Goals} 
Applying our framework to the practical design of WebSeek revealed several intricacies of real-world, web-based sensemaking. To address these, we distilled our learnings into four core design goals. These goals guided the system's implementation and serve as prescriptive principles for future mixed-initiative data tools.

\textbf{DG1: Infer User Intent from Holistic Interaction Context.}
To provide truly helpful assistance, the AI must move beyond simple triggers and infer user intent from a rich, holistic view of the user's context. A single interaction cue, such as selecting a table, is an ambiguous signal on its own. We posit that the system must synthesize information from the full spectrum of available context, including: the current system state (e.g., the artifacts on the canvas), the user's recent interaction history, the specific webpage and system view the user is focused on, and the conversational history with the AI. By integrating these multiple signals, the system can build a more accurate model of the user's immediate goal, forming the foundation for more relevant and timely assistance.

\textbf{DG2: Prioritize a Core Set of Complementary Interaction Modalities around Data Instances.}
A powerful system must provide users with effective ways to express their intent, but an overabundance of interaction modalities can lead to a cumbersome and confusing experience. During our iterative design process, we explored various modalities, including code-based panels for instance state management and the creation of sketch instances on the canvas for visual prompting. However, user feedback revealed that this overwhelming multimodal richness introduced significant cognitive overhead. Therefore, our design goal became to balance expressiveness with learnability~\cite{l2024learnable} by deliberately prioritizing a core set of highly complementary modalities. 

Given that WebSeek is mainly targeted at data analysts and everyday web users who may not necessiarly possess a strong technical background, we eventually opted for a design that prioritized \textit{direct manipulation} and \textit{textual prompts} due to their intuitiveness and wide application scope.  Direct manipulation can not only be used to directly construct instances but also serve as demonstrations or example results for AI inference. Textual prompts complement this by providing easily accessible and efficient support for diverse tasks. 
This focused combination provides wide coverage for web data tasks without overwhelming the user.

\begin{figure*}[t]
  \centering
  \includegraphics[width=\linewidth]{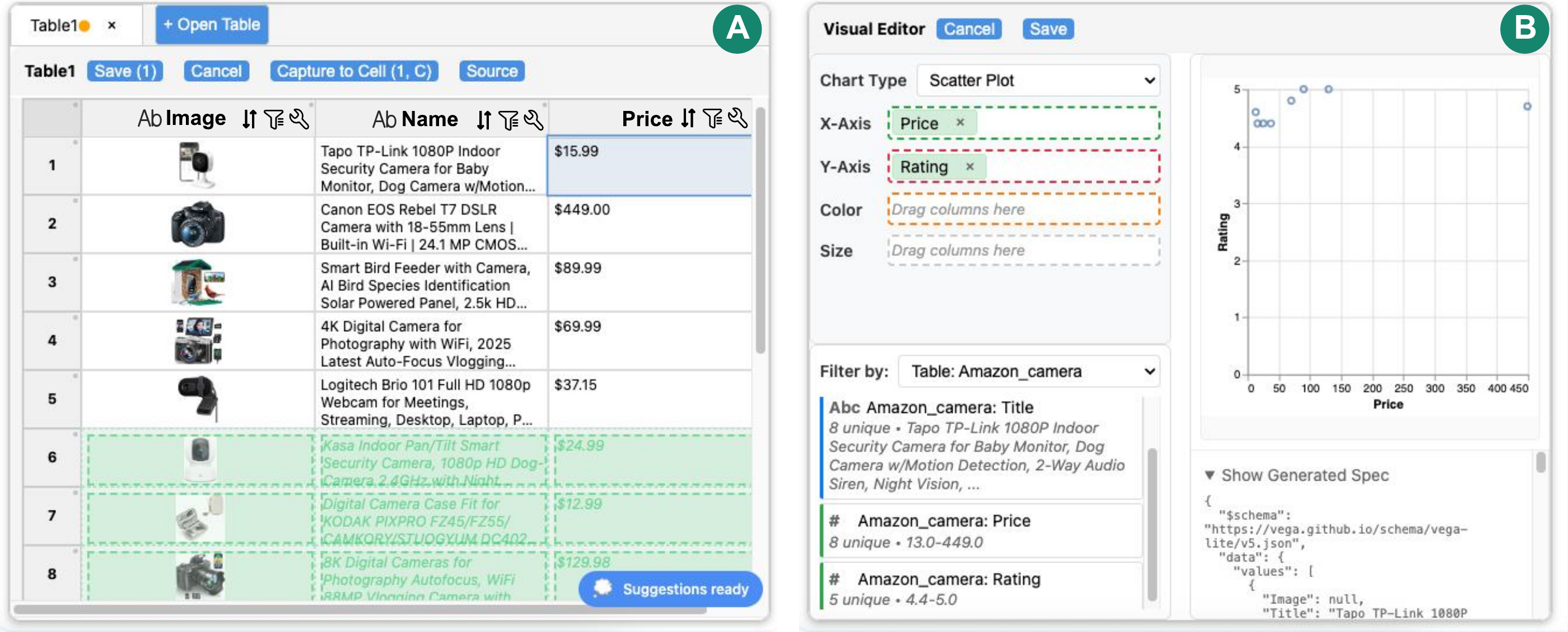}
  \caption{The table editor (A) and visualization editor (B) in WebSeek. In the table editor, an in-situ suggestion is provided suggesting the completion of the next few rows in green after users fill the initial ones. In the visualization editor, users may select a chart type and drag data attributes to the shelves (x-axis, y-axis, color, and size) to create a visualization.}
  \Description{In the table editor on the left, there is a classical spreadsheet interface which includes three columns currently: Image, Name, and Price. The initial five rows have been filled, and an in-situ suggestion is displayed, completing the next few rows with those cells colored in green. In the Visualization Editor on the right, there are several shelves on the left of it (x-axis, y-axis, color, and size), where users can drag data attributes below to these shelves to create visualizations. The visualizations will be rendered on the right which users can interact with.}
  \label{fig:editor}
\end{figure*}

\textbf{DG3: Surface Composite Suggestions as Transparent, Multi-Step Plans.}
In the framework, we keep most example AI proactive actions atomic for generality. However,  the suggestions tend to have many prerequisite steps in practice. For instance, suggesting a visualization of price data (a macro-goal) is useless if the Price column is still formatted as text. This necessitates the use of \textit{Composite Suggestions}, where the AI presents a complete, multi-step plan as a single, cohesive unit. Our design goal is to make these plans transparent to keep users aware of the AI's entire line of reasoning (e.g., ``Step 1: Convert 'Price' to a number. Step 2: Create a bar chart.''). This approach respects user agency by allowing them to understand and approve the full scope of an action.

\textbf{DG4: Ground AI Actions in a Reliable, Tool-Based Execution Architecture.}
To ensure that accepted suggestions are executed predictably and correctly, the system's actions must be grounded in deterministic operations. While LLMs are excellent at high-level planning, allowing them to directly generate the final data state is unreliable and prone to hallucination. Inspired by the recent advances in tool-based AI agents, our final design goal is to define a library of concrete, reliable tools (e.g., \texttt{tableSort()}, \texttt{convertColumnType()}) that the AI can call. In this model, the LLM acts as a planner, outputting a sequence of tool calls (as established in DG3).The WebSeek system then executes these calls using its own well-tested code. This tool-based architecture ensures that all AI-driven actions are as reliable, predictable, and reversible as manual user manipulations.

\textbf{DG5: Adhere to the ``User-Centered AI'' philosophy rather than ``AI-Centered User''~\cite{narechania2025agentic}.} This principle mandates that all AI assistance must respect the user's workflow, focus, and ultimate authority over the workspace. To operationalize this, we implemented several key policies to ensure the AI acts as a deferential assistant rather than an intrusive agent:

\begin{enumerate}[label={\alph*)}]
    \item \textbf{Conflict Avoidance:} To prevent interference with the user's actions, in-situ suggestions are immediately aborted if the user is actively typing or interacting with the target element. While merging the earlier AI suggestions with later user interactions might be feasible, we learned through our attempts the intricacies within such a method that might be out of the paper's scope. We hence prioritized a clean, non-conflicting user experience by having the AI yield to direct user input.
    \item \textbf{Respect for User Focus:} To maintain awareness and control, the AI is constrained from modifying any data artifacts that are not currently within the user's viewport. If a necessary change affects an off-screen element, the system must first prompt for user permission and offer to navigate them to the relevant context.
    \item \textbf{System State Transparency:} The AI's operational status is always visible. A status indicator informs the user when the AI is processing information, and a subtle notification appears when new suggestions are ready, ensuring the user is never left guessing.
    \item \textbf{Maintaining Relevance:} Suggestions are treated as ephemeral and context-dependent. The system automatically clears older peripheral suggestions when it detects a strong signal of an intent change, such as the user issuing a new command in the chat view, ensuring the AI's guidance remains timely and relevant to the user's current goal.
\end{enumerate}




\section{WebSeek}

\begin{figure*}[t]
  \centering
  \includegraphics[width=\linewidth]{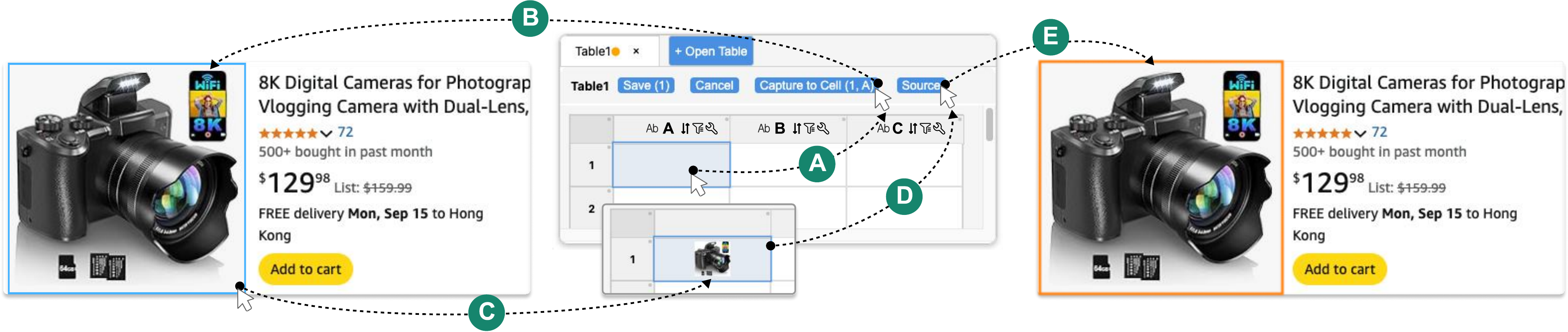}
  \caption{Illustration of the data capture and source tracing interactions. (A) The user click on an empty cell and clicks on the ``capture'' button. (B) The user enters the selection mode and is enabled to capture DOM elements. (C) The captured image is saved in the cell. (D) The ``source'' button is clicked to locate the data source on the web. (E) The browser is automatically navigated to the source page with the source DOM highlighted. }
  \Description{In Step B, the users clicks on an image of a camera on the webpage to capture the image into the cell. In Step E, the user is navigated to the camera item with the DOM of the image highlighted on the webpage.}
  \label{fig:capture}
\end{figure*}

In this section, we present the detailed design and implementation of WebSeek. To support seamless data analysis and decision making in a uniform environment, WebSeek is developed as a browser extension that can be opened as a side panel based on the WXT framework\footnote{https://wxt.dev/}.

\subsection{Interface design}

\autoref{fig:interface} show the interface of WebSeek. It consists of two views: an AI guidance view (A) and a workspace view (B). In the AI guidance view, the users can view the peripheral proactive AI guidance  (A1) and chat with the LLM to communicate or modify the state of the data instances in the workspace. The workspace view (B) is a canvas where users are allowed to create data table (B1-B4) or visualization instances (B5). 

\subsubsection{Manual Instance Creation and Editing}
Informed by DG2, WebSeek enables users to directly manipulate the instances for fine-grained control. Specifically, it allows users to manually create and edit table or visualization instances through dedicated editors.

The table instance editor is shown in \autoref{fig:editor}. It supports a rich set of common interactions in spreadsheet tools like Excel. For instance, users may double-click on cells to edit, write formulas and do flashfilling, copy and paste cells, select cells, rows, and columns and delete or add them. In additional to basic ones, WebSeek supports advanced data transformation operations based on the space from Kandel et al.~\cite{wrangler}, including \textit{map transforms}, \textit{table joins}, \textit{reshape transforms}, \textit{positional transforms}, and \textit{aggregation}. Frequently used actions, such as sorting, filtering, and changing column types, are available via icons next to column headers. Less frequently used operations, like formatting, splitting, cutting \& extracting, and character replacement, are grouped in the panel accessed via the tool icon on the far right.

Additionally, to accommodate the tasks and features in our framework, two interactions are designed to assist users with data extraction and source tracing on the web. As depicted in \autoref{fig:capture}, users can click on the ``capture'' button to enter the selection mode and freely capture DOM elements. Users can also navigate to the data source at any time by clicking on the ``source'' button.

Meanwhile, the visualization editor shown in \autoref{fig:editor} enables shelf-based creation and editing of visualizations. Users may drag the data attributes from the bottom right to the shelves to construct Vega-Lite\cite{satyanarayan2016vega} specifications, which will be automatically augmented with interactive features (zooming and panning, tooltips, and filtering) before generation.

\begin{table*}[t]
\centering
\caption{Tools provided by WebSeek organized by task category.}
\label{tab:tools}
\footnotesize
\renewcommand{\arraystretch}{1.2}
\begin{tabular}{>{\raggedright}p{1.6cm}>{\raggedright}p{2.6cm}>{\raggedright\arraybackslash}p{5.5cm}>{\raggedright\arraybackslash}p{5.3cm}}
\toprule
\textbf{Category} & \textbf{Feature} & \textbf{Tool} & \textbf{Description} \\
\midrule
\addlinespace[2pt]
\multirow{1}{1.8cm}{\textbf{Discovery}} 
& Suggest useful webpages 
& \texttt{\small openPage(url, description, ...)} 
& Opens a webpage in a new browser tab for suggesting useful data sources related to the user's current work \\[4pt]
\midrule
\addlinespace[2pt]
\multirow{7}{1.8cm}{\textbf{Data Extraction \& Wrangling}} 
& Element selection 
& \texttt{\small selectElements(selector, pageUrl, ...)} 
& Automatically identifies and selects relevant DOM elements on webpages for data extraction \\[2pt]
\cmidrule{2-4}
& Schema inference 
& \texttt{\small inferSchema(pageUrl, targetElement, ...)} 
& Analyzes webpage structure and infers data schema for structured extraction from tables and lists \\[2pt]
\cmidrule{2-4}
& Batch extraction of list/table entries 
& \texttt{\small extractBatch(pageUrl, pattern, maxItems, ...)} 
& Extracts multiple similar data entries from web pages in batch operations using pattern recognition \\[2pt]
\cmidrule{2-4}
& Row/Column Autocomplete 
& \texttt{\small updateInstance(instanceId, newInstance)} 
& Updates the specified instance with new values \\[2pt]
\cmidrule{2-4}
& Computed/derived columns 
& \texttt{\small addComputedColumn(instanceId, formula, newColumnName, ...)} 
& Creates new columns based on formulas or calculations derived from existing columns \\[2pt]
\cmidrule{2-4}
& Sorting and Filtering 
& \texttt{\small tableSort(instanceId, columnName, order, ...)} 
& \multirow{2}{5.7cm}{Applies sorting to organize data and filtering to show/hide rows based on criteria for data refinement} \\
& & \texttt{\small tableFilter(instanceId, conditions, ...)} 
& \\[2pt]
\cmidrule{2-4}
& Joining tables 
& \texttt{\small mergeInstances(sourceInstanceIds, mergeStrategy, joinColumns, ...)} 
& Combines multiple table instances using union, inner join, left join, or right join operations with flexible column mapping \\[4pt]
\midrule
\addlinespace[2pt]
\multirow{4}{1.8cm}{\textbf{Data Profiling \& Cleaning}} 
& Entity Resolution \& Normalization 
& \texttt{\small renameColumn(instanceId, oldColumnName, newColumnName)} 
& \multirow{2}{5.7cm}{Renames columns and standardizes entity names across datasets} \\
& & \texttt{\small formatColumn(instanceId, oldColumnName, formatPattern, ...)} 
& \\[2pt]
\cmidrule{2-4}
& Remove extraneous characters 
& \texttt{\small searchAndReplace(instanceId, searchPattern, replaceWith, ...)} 
& Performs find and replace operations across text-based instances with support for regex patterns for advanced text manipulation \\[2pt]
\cmidrule{2-4}
& Data type correction 
& \texttt{\small convertColumnType(instanceId, columnName, targetType, ...)} 
& Converts table columns between data types (e.g., converting currency strings to numbers) with optional cleaning patterns \\[2pt]
\cmidrule{2-4}
& Fill missing values 
& \texttt{\small fillMissingValues(instanceId, columnName, strategy, ...)} 
& Fills missing or empty cells using strategies like mean, median, mode, interpolation, or constant values \\[4pt]
\midrule
\addlinespace[2pt]
\multirow{3}{1.8cm}{\textbf{Data Modeling \& Visualization}} 
& Auto-generate visualizations 
& \multirow{2}{6cm}{\texttt{createVisualization(sourceInstanceId, chartType, xAxis, yAxis, ...)}} 
& \multirow{2}{5.5cm}{Creates new visualization instances (bar, line, scatterplot, and histograms).} \\
\cmidrule{2-2}
& Suggest alternative charts 
& 
& \\[2pt]
\cmidrule{2-4}
& Interactive filtering/highlighting 
& \texttt{\small tableFilter(instanceId, conditions, operator, ...)} 
& Applies interactive filtering and creates dynamic highlighting systems for data exploration \\[2pt]
\bottomrule
\end{tabular}
\end{table*}

\subsubsection{Tool Design and Guidance}

Informed by DG4, WebSeek provides a comprehensive set of tools for each task category and feature in our framework. \autoref{tab:tools} presents a detailed overview of all tools implemented in WebSeek, including their functionality and corresponding API signatures. These tool are used to power three types of guidance in WebSeek:

\begin{itemize}
    \item \textbf{In-situ completion}: Micro suggestions are generated only when users are in an editor. They are presented as in-situ completions, with the different cells colored as green or red to indicate addition and deletion. Distinctions will be highlighted in cells that need to be modified. \autoref{fig:editor}(A) shows an example of completing multiple table rows.

    \item \textbf{Peripheral suggestions}: Macro suggestions are generated when users are idle for over 5 seconds regardless of the user's current view. A description and a tool sequence will be shown to users alongside the feature (DG3).

    \item \textbf{Chat-based guidance}: Users can reactively seek AI assistance in the chat view through textual prompts. They may also use ``@'' to refer to existing instances in the workspace through their names, where a selector for auto-completing names will be shown when the character is added to the input box.  
\end{itemize}

In our implementation, all guidance are generated by Gemini-2.5-Flash with carefully engineered prompts. Each prompt includes five types of context (DG1): 

\begin{itemize}
    \item \textbf{HTML Context:} The raw HTML of all opened webpages, captured by WebSeek when it is opened. When the content of the current webpage changes or a new webpage is opened, WebSeek will automatically detect such changes and capture the new version. All HTML context will be saved on the backend and fetched when needed.

    \item \textbf{Instance Context:} The underlying data or specifications of all instances in the instance view.

    \item \textbf{User's focus:} The current view the user is in as well as the current active tab in the browser (DG5).

    \item \textbf{Conversation history:} All conversations in the chat view.

    \item \textbf{Interaction history:} The logs of the latest 15 major user interactions (i.e., trivial interactions without user intent information like moving instances will not be counted).
\end{itemize}

Admittedly, incorporating all contexts within a LLM call may lead to context explosion as users continues to use the system. To address this issue, we tried to design an agentic AI using open-source agent frameworks in our design process, where we supported dynamic context fetching by defining a function ``\texttt{fetchContext()}'' in the frontend for the agent to call with. However, we later learned that the context would often be fetched for multiple times during an agent call, leading to prohibitively high time cost. We envision more efficient generative AI models and agent frameworks to address this issue and we leave this for future work.

In case multiple suggestions need to be presented to the user, we instructed the LLM to return a confidence score along with the suggestion. The ones with higher confidence score will be prioritized in the display order. 

\begin{figure*}[t]
  \centering
  \includegraphics[width=\linewidth]{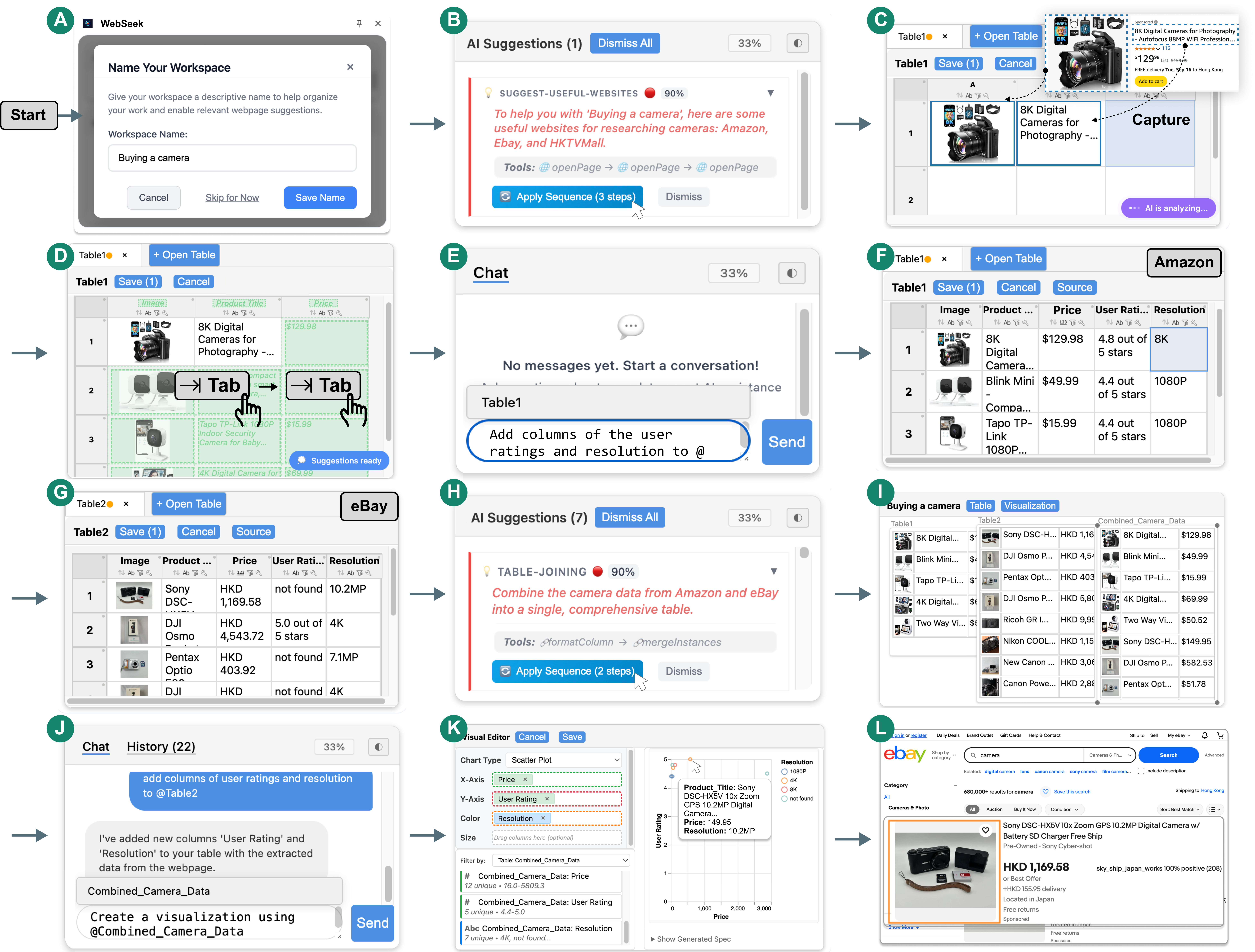}
  \caption{An illustration of the usage scenario. (A) The user names the workspace before entering the interface. (B) A peripheral proactive AI suggestion is generated. (C) The user captures data from Amazon into cells. (D) An in-situ proactive AI suggestion is generated twice, and the user taps on the \textit{tab} key to accept it. The user repeats this to get the complete table. (E) The user chats to add new columns. (F) The extracted table from Amazon. (G) The user opens eBay and a new table is extracted simiarly. (H) An proactive AI suggestion for joining two tables is generated. (I) The results after the table join. (J) The user chats to create a visualization. (K) The user checks the generated visualization and navigates to the data source of a chosen item. (L) The item's source highlighted on eBay.}
  \Description{In Step (A), the user enters the workspace name "Buying a camera". In Step (B), the AI suggestions view displays a suggestion with type "suggest-useful-websites": \textit{To help you with 'Buying a camera', here are some useful websites for researching cameras: Amazon, Ebay, and HKTVMall.} It also suggests three \textit{openPage} actions in sequence. In Step (C), the user captures the image and the title of a camera item from the webpage. In Step (D), the in-situ suggestion automatically fills the initial few rows after the user fills the first two cells of the table. In Step (E), the user types a prompt in the Chat view: "Add columns of the user ratings and resolution to @". At this time, an auto-completion dropdown appears, showing "Table1",  the ID of an existing table. (F) The completed table from Amazon, including five columns: Image, Product Title, Price, User Rating, and Resolution. (G) The completed table from eBay with the same schema with the Amazon table. (H) The AI suggestions view displays several suggestions, and one of them is "Table-joining: Combine the camera data from Amazon and eBay into a single, comprehensive table", with two actions in sequence: "formatColumn" and "mergeInstances". (I) The instance view after applying the suggestion, including three tables: the Amazon table ("Table1"), the eBay table ("Table2"), and the merged table ("Combined\_Camera\_Data"). (J) The users enters the prompt "Create a visualization using @Combined\_Camera\_Data" in the Chat view. (K) The AI suggests a scatterplot with "Price" mapped to the x-axis, "User Rating" mapped to the y-axis, and "Resolution" mapped to the color. (L) The user right-clicks on an interested point in the plot and navigates to the item on the source webpage, eBay.}
  \label{fig:scenario}
\end{figure*}

\subsection{Usage scenario}

We now go through the workflow of WebSeek through a scenario of product research. Imagine a marketer, Peter, who needs to research the best camera to buy from different e-commerce platforms for his company.

Peter opens WebSeek and starts by creating a workspace called ``Buying a camera'' (\autoref{fig:scenario}(A)). Based on this initial context, the system's proactive AI immediately provides a peripheral suggestion with links to relevant e-commerce websites like Amazon and eBay, which Peter can use to kickstart his research (\autoref{fig:scenario}(B)). Peter opens the Amazon link and begins his data collection. He uses WebSeek's direct manipulation feature to capture the image and product title of the first camera into a new table artifact on his canvas (\autoref{fig:scenario}(C)).

After Peter captures the second item in the list, the AI detects a repetitive pattern and provides an in-situ suggestion to autocomplete a few rows with the corresponding data and meanwhile update the column names. Peter accepts this by simply pressing the Tab key, rapidly accelerating his data extraction process (\autoref{fig:scenario}(D)). Detecting the acceptance, the AI continues to suggest a few more rows and Peter presses the Tab key again to get more products. Realizing he missed some key attributes, Peter uses the reactive chat interface to issue a command:  ``\textit{Add columns of the user ratings and resolution to @Table1}'' (\autoref{fig:scenario}(E)). The system executes this, enriching his table with the new data from the Amazon page (\autoref{fig:scenario}(F)).

Next, Peter navigates to eBay to gather more data and repeats a similar extraction process to create a second table (\autoref{fig:scenario}(G)). With two distinct tables now on his canvas, the AI identifies a strategic opportunity. It generates a new macro-suggestion in the peripheral panel, proposing to join the two tables into a single, comprehensive dataset. Meanwhile, it detects the inconsistency of price currency and includes the formatting operation as a prerequisite step in the tool-based plan (\texttt{formatColumn → mergeInstances}) it suggests (\autoref{fig:scenario}(H)). Peter accepts, and the two tables are seamlessly merged into one unified view (\autoref{fig:scenario}(I)).

To better understand the relationship between price, rating, and resolution in his combined dataset, Peter again uses the chat to ask the system to ``\textit{Create a visualization using @Combined\_Camera\_Data}'' (\autoref{fig:scenario}(J)). WebSeek generates a scatterplot and opens the visual editor, where Peter can refine the encodings (\autoref{fig:scenario}(K)). While exploring the chart, he notices an interesting data point with the highest user rating and medium price. He hovers over it to see its details and decides to investigate further. By clicking on the data point, WebSeek demonstrates its data-source traceability, instantly navigating him back to the original eBay product listing and highlighting the corresponding item on the webpage (\autoref{fig:scenario}(L)). He eventually adds this item to the shopping cart. This seamless cycle of data gathering, AI-assisted wrangling, visualization, and source verification allows Peter to conduct his complex research within a single, integrated environment.

\section{Technical Evaluation}
To demonstrate the robustness of WebSeek's guidance, we conducted a technical evaluation assessing the system's accuracy in generating in-situ and peripheral suggestions across diverse web domains and tasks of varying levels of difficulty. Our methodology was inspired by ProactiveVA~\cite{zhao2025proactiveva}'s study on LLM-generated proactive guidance for visual analytics systems, where the performance of the guidance generation algorithm was assessed. With a similar study goal, we adapted its methodology to the context of web-based decision-making.

\begin{table*}[t]
\centering
\caption{Average count of generated suggestions, average latency, and accuracy by task difficulty and guidance type in the technical evaluation.}
\label{tab:latency_accuracy}
\begin{tabular}{l c c c c c c c c}
    \toprule
    Difficulty & Count
               & \multicolumn{2}{c}{Average Suggestion Count}
               & \multicolumn{2}{c}{Average Latency (s)}
               & \multicolumn{2}{c}{Accuracy} \\
    \cmidrule(lr){3-4} \cmidrule(lr){5-6} \cmidrule(lr){7-8}
               & 
               & In-situ & Peripheral
               & In-situ & Peripheral
               & In-situ & Peripheral \\
    \midrule
    Easy   & 20 &  9.0 & 14.7 & 11.44 & 3.35 & 100.0\% & 93.8\% \\
    Medium & 20 & 17.3 & 38.3 & 13.01 & 3.39 & 98.7\% & 92.9\% \\
    Hard   & 10 & 19.5 & 46.7 & 19.05 & 3.55 & 100.0\% & 80.0\% \\
    \bottomrule
  \end{tabular}
\end{table*}

\subsection{Methodology}
\subsubsection{Tasks}
Considering there are few existing benchmarks targeting web-based data wrangling or data-driven decision-making tasks, we created a custom benchmark of 50 tasks, following the approach used in the evaluation of the ProactiveVA framework~\cite{zhao2025proactiveva}. The tasks were designed to encompass diverse activities, including data extraction, wrangling, and visualization. A single task could integrate multiple such activities. Each task comprised a statement describing the task goal and one or more input webpages. The inputs were drawn from a set of 22 webpage snapshots\footnote{We used snapshots rather than live pages to ensure a controlled, reproducible study.} spanning 17 distinct domains (e.g., finance, sports, movies). 
Each resultant webpage included between 10 and 40 data records in the main body.
To ensure a rigorous evaluation of the system's robustness against real-world data quality issues, we manually infused realistic variations in data schema and quality (e.g., inconsistent entity formats and missing values) into these snapshots.
We also manually validated these tasks to ensure coverage across the entire feature space (\autoref{tab:space}).
These snapshots and tasks were generated by prompting Claude Sonnet 4.5 and GPT-4 Turbo respectively, and the corresponding prompts are available in Appendix.

Additionally, to evaluate WebSeek across a range of task complexities, we adopted a difficulty metric inspired by prior work~\cite{yu2018spider,narechania2021diy}. Specifically, we determined a task’s difficulty level based on the number of criteria it meets out of the following list: (a) multiple source webpages are required, (b) more than five data transformation operations are needed, and (c) data visualization is required. Tasks meeting 0 criteria are labeled Easy, 1–2 criteria as Medium, and all 3 criteria as Hard. This scheme was leveraged during the task generation and eventually, the benchmark comprised 20 Easy, 20 Medium, and 10 Hard tasks.

\subsubsection{Procedure} 
We used an efficient reasoning LLM, Grok Code Fast 1, as a virtual user that simulates real users, while calling WebSeek to generate guidance throughout the task. 
The model received real-time system state, including HTML context, data instances and prior guidance, to determine the next action, which one author then performed manually in WebSeek as it requires human browser interaction. If the model generated a wrong action that was impossible to perform in WebSeek, it would be asked to regenerate the action with a warning ``\texttt{[INVALID ACTION]}'' added to the interaction logs of the new system state.
We iterated this process until task completion, logging latency, the guidance description, and system states before/after guidance whenever a guidance was applied. This process resulted in a set of 107 guidance items. Similar to the ProactiveVA~\cite{zhao2025proactiveva} study, two data analysts with over two years of data analysis experience then evaluated the accuracy of each applied guidance, assessing if the state changes aligned with the guidance description and labeling it as \textit{correct}, \textit{incorrect}, or \textit{not sure}. In the following analysis, we excluded the one item on which at least one annotator labeled as \textit{not sure}, and for the three items where the labelers disagreed, we assigned a partial credit of 0.5 correct each. 

\subsection{Results}
\autoref{tab:latency_accuracy} shows the results of our technical evaluation. On average, WebSeek was able to generate different types of guidance within 20 seconds of interactions. In addition, the system maintained a high level of accuracy, achieving an overall rate of 97.2\%. Remaining failure cases mainly stem from the failure of the model to determine the correct data type, mistakes in the format of tool calls, and  hallucinations. However, the virtual user still managed to finish all tasks, as WebSeek handled such cases by regenerating the problematic suggestion and allowing users to interact alternatively (e.g., seeking reactive guidance or direct manipulation) to finish the tasks.

\section{User Study}


To understand how WebSeek enables transparent human–AI collaboration for web data analysis and decision making, we conducted an exploratory user study using it as a design probe. Specifically, we studied how participants perceive and utilize WebSeek's core features including proactive and reactive AI guidance and interactions with tangible data artifacts using different interaction modalities. We adopted this methodology for two main reasons: first, since WebSeek incorporates novel design heuristics, an exploratory study can help us capture and understand users' nuanced behaviors and perceptions. Second, to the best of our knowledge, WebSeek is the first human-centered browser extension built for human-AI collaborative data analysis on the web, so there are no suitable baselines for comparison yet. Even so, we included confirmatory measures to rigorously evaluate its overall usability and effectiveness.

\begin{figure*}[t]
  \centering
\includegraphics[width=0.8\linewidth]{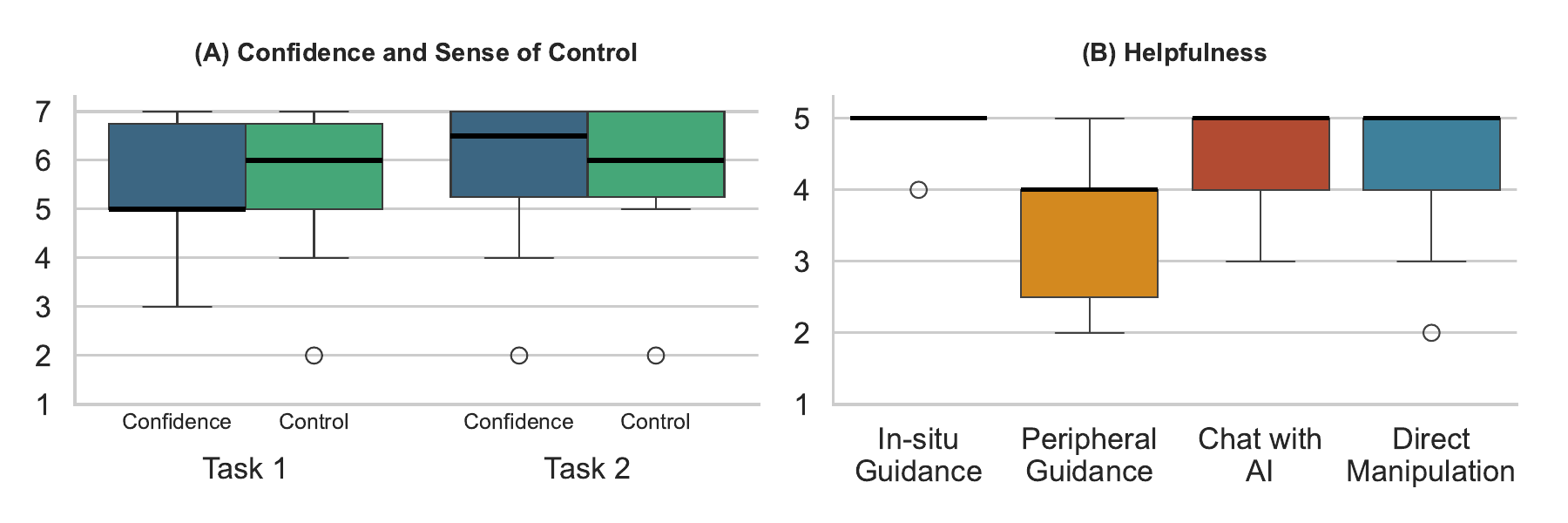}
  \caption{Participants' ratings on (A) their perceived confidence on the final results and sense of control during the process, and (B) the perceived helpfulness of each system feature.}
  \Description{The figure contains two side-by-side panels. The left one measures users' confidence and sense of control in both tasks on a 7-point scale, and the right one  measures participant responses on a 5‑point scale for four different features: In‑situ Guidance, Peripheral Guidance, Chat with AI, and Direct Manipulation.}
  \label{fig:module_overview}
\end{figure*}

\begin{figure*}[t]
  \centering
\includegraphics[width=\linewidth]{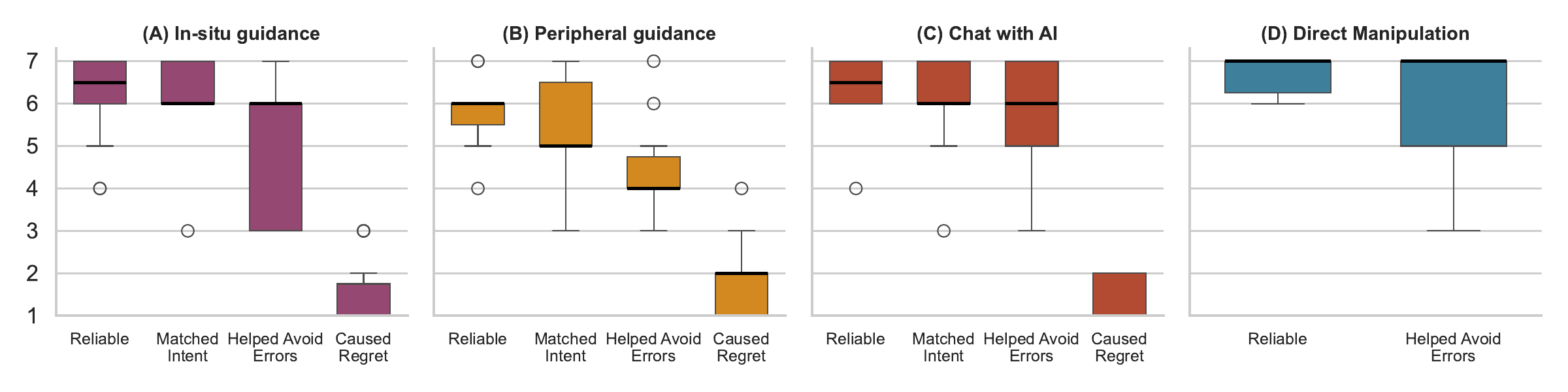}
  \caption{Participants' detailed ratings on the four features of the system regarding 1) whether the guidance/feature was reliable, 2) whether the guidance matched the user's intent, 3) whether the guidance/feature helped avoid errors, and 4) whether the user regretted following the guidance.}
  \Description{The figure shows users' detailed ratings on the above four dimensions for the four features in the four subplots. It can be shown that users believed the in-situ guidance, chatting with AI, and direct manipulation are more reliable and error-reducing than peripheral guidance. There are similar patterns in terms of the intent matching and casued regret for in-situ guidance and chatting with AI compared with peripheral guidance.}
  \label{fig:module_detail}
\end{figure*}

\subsection{Study Design}
\subsubsection{Participants} We recruited 15 participants (9 male, 6 female), including 2 undergraduates, 6 Master's, and 7 Ph.D. students recruited from two academic institutions. The majors of these participants range from Computer Science (6), Medicine (2), Electrical Engineering (2), Artificial Intelligence (1), Chemistry (1), Mechanical Engineering (1), Energy Engineering (1), Design (1). Participants had experience in using visual data analytics tools like Excel and Tableau or programming for data analysis, with an average self-reported score of 3.67 on a 5-point Likert scale (SD=0.724).

\subsubsection{Settings} Participants were allowed to either use their own PC if it is at least 14-inch and 1080P, or work on a monitor provided by us (24-inch, 1080P). All of them worked on Chrome (minimum version: 130.x) or Microsoft Edges (minimum version: 140.x).

\subsubsection{Tasks} Two tasks were involved in this study. For diverse coverage, the first task was in an artificial setting with an existing ground truth, while the second task was a high-fidelity real-world scenario with an open-ended nature adapted from prior literature~\cite{malleable}. We made sure these tasks covered the task categories and features mentioned in the framework mentioned in Section 3.

\begin{itemize}
    \item \textbf{Task 1: Fact-checking a news story.} In this task, participants were asked to examine the following news story: ``\textit{In Olympic Games, simple medal tables are misleading. The most impressive Olympic stories are the outliers—the countries that vastly overperform relative to their population. As a result, economic might (GDP) is a better indicator of a nation’s sports power than population.}'' Considering the random size and tidiness of arbitrary real-world data sources, we prepared two webpages as data sources for participants to control the difficulty and efficiency of the study. Each webpage contains a data table of 30 rows. The first table is a relational table that includes the fields \textit{country} and \textit{gold}, \textit{silver}, \textit{bronze}, \textit{total} medals in an Olympic Games. The second table is a relational table that includes the fields \textit{country}, \textit{GDP}, and \textit{population} for the same year. To mimic real-world web data, both tables contain several data quality issues such as inconsistent entities, missing values, and redundant special characters in columns, requiring participants to carefully clean the data either manually or by working with the AI. 

    \item \textbf{Task 2: Product Comparison.} Participants were tasked with a scenario where they were setting up a home office and needed to buy a monitor on e-shopping platforms like Amazon and Ebay with a budget of HKD 1000. They were asked to look for the sweet spot between different criteria such as price, user review and/or rating, resolution, and so on. Participants could also define their own selection criteria and were required to justify their choice using supporting data. They were encouraged to behave as they normally would when shopping online, and there was no limit on the number of platforms they could consult.
\end{itemize}

\subsubsection{Procedure} After acquiring the consent from participants, participants were asked to fill in a demographic survey. Next, each participant received a brief tutorial on the system, followed by a 10-minute warm-up session in which they tried every system feature. They then worked on the two tasks, with a 1-minute break between tasks. The session ended with a post-study questionnaire and a semi-structured interview. 
Each study lasted approximately 90 minutes, and participants were compensated HKD 90. The study protocol has been approved by the IRB committee of our institution.

\subsubsection{Measure} Participants were asked to rate the usability of WebSeek using the System Usability Scale (SUS)~\cite{brooke1996sus} and the helpfulness of different components of the system. 
To study the impact of AI guidance in detail, we asked targeted questions probing participants’ perceptions of the guidance along dimensions such as intent alignment, perceived reliability, usefulness for reducing errors, and the sense of regret caused. For the manual  functionalities, participants answered two questions about realiability and error-reducing.
We also collected the interaction logs while recording their screen throughout the study.

\subsection{Findings}
\subsubsection{Finish Rate, Time, and User ratings.} All participants successfully finished both tasks. In Task 1 (fact-checking), the ground truth of the story should be true, supported by Pearson correlations between relevant data attributes (r = 0.91 for \textit{GDP} vs. \textit{Total Medals} and r = 0.59 for \textit{Population} vs. \textit{Total Medals}. This fact remains true if the participant chose other medal types).
We discovered that 12 out of 15 participants reached a conclusion that matched the ground truth. Of the three failures, P1 produced charts from the wrong dataset because of an earlier data-cleaning error that corrupts a data column.
P10 and P15 created the correct charts but interpreted them differently: P10 overly focused on a few outliers and misjudged the trends, while P15 believed that there were no significant differences between the two indicators. In Task 2, participants used diverse criteria to determine the product to buy, including the user rating (13/15),  size (6/15), refresh rate (3/15), sales (3/15), resolution (2/15), the number of users rating the product (2/15), and platform (1/15). We also noticed two participants who defined a custom score for each product using a self-designed formula, and one participant managed to compared  products from multiple platforms. Overall, participants spent 14.60 minutes (SD=4.47) and 12.65 minutes (SD=6.58) on the two tasks, respectively.

In addition, participants were generally positive on WebSeek's usability, giving an average SUS score of 73.11/100 (SD=14.50). Meanwhile, as depicted in \autoref{fig:module_overview}(A), they also reported for both tasks a high degree of confidence (M=5.33/7, SD=1.59 for Task 1; M=5.67/7, SD=1.54 for Task 2) and sense of control (M=5.80/7, SD=1.08 for Task 1; M=5.53/7, SD=1.41 for Task 2). Regarding the helpfulness of the features (\autoref{fig:module_overview}(B)), participants generally spoke highly of  in-situ guidance (M=4.93/5, SD=0.26), chatting with AI (M=4.53/5, SD=0.74), and direct manipulation (M=4.40/5, SD=0.99), while the ratings on the peripheral guidance are mixed and notably lower (M=3.42/5, SD=1.08). A similar pattern exists in the detailed ratings on these features (\autoref{fig:module_detail}): generally, participants believed the top three features were reliable and error-reducing, with the guidance well-aligned with user intent and causing little regret. Peripheral guidance, however, was less consistently seen as helpful though viewed as positive on average.

\subsubsection{User Behavior} 

\begin{figure*}[t]
  \centering
\includegraphics[width=\linewidth]{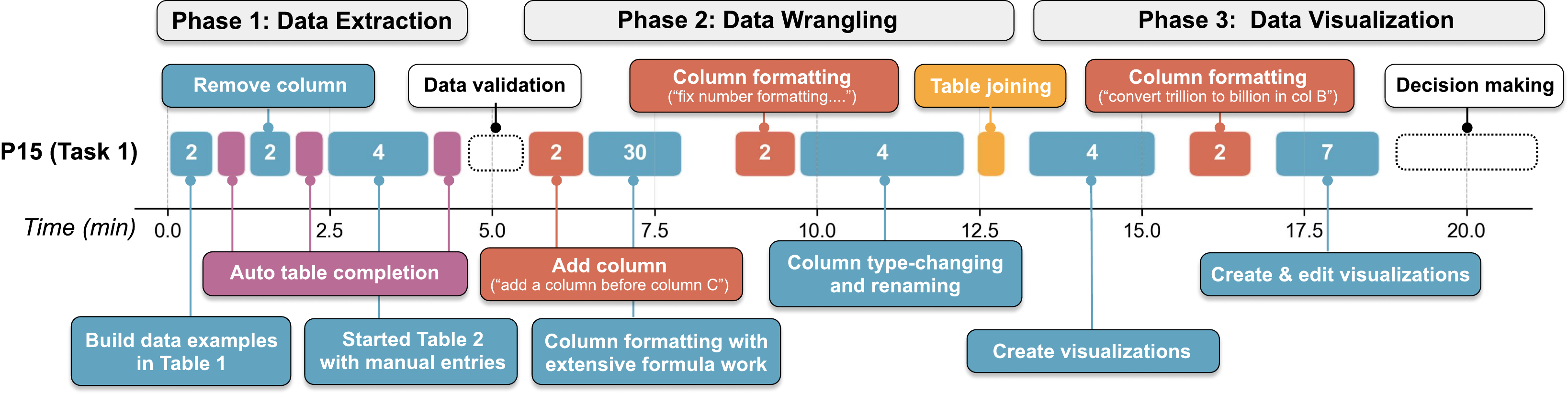}
  \caption{A visualization of P15's workflow on Task 1. The color encoding is the same as previous figures (i.e., purple: in-situ suggestions; yellow: peripheral suggestions; orange: chatting with AI; blue: manual interactions). Consecutive events of the same category close to each  other are merged into blocks, and the numbers on the blocks indicate the number of events merged (no merging if not shown).}
  \Description{The workflows shows three main stages: Data Extraction (0-5min approximately), Data Wrangling (5-13min approximately), and Data Visualization (13-23min approximately). In the Data Extraction Stage, the user built data examples in Table 1, applied in-situ table completion, removed columns, applied in-situ table completion, started Table 2 with manual entries, applied in-situ table completion, and did data validation. In the Data Wrangling stage, the user added a column by chatting with AI, formatted columns with extensive formula work, chatted with AI for column formatting, changed column types and renamed columns, and joined tables through applying proactive AI suggestions. In the Data Visualization stage, the user created visualizations, formatted columns by chatting with AI, edited visualizations, and did decision making. }
  \label{fig:p15a}
\end{figure*}

We analyzed the participants' behavior patterns during the tasks by analyzing the log data with the following steps. First, we filtered out trivial interactions like open/close editors and repositioning instances, focusing on actions that were crucial to the task outcome such as editing instances and applying suggestions. The result events are classified into four categories: \textit{in-situ guidance}, \textit{peripheral guidance}, \textit{chatting with AI}, and \textit{direct manipulation}. 
Second, we merged consecutive interactions of the same category if their time gap is within a predefined threshold (1.5 min in our analysis) to reveal patterns from the massive logs. The merged result could be heuristically viewed as an active period where users were devoted to an interaction modality.
Finally, we visualized the interaction timelines for each participant, inspected the patterns, and annotated interesting  individual trails. In total, we collected logs of 29 valid sessions (15 participants $\times$ 2 tasks; the logs of P5 on Task 1 was missing due to some technical issues). We will discuss below our major findings with a few selected examples, and the complete visualizations are included in the supplemental materials.

\textbf{Statistics of the task completion process.}
Overall, participants spent 67.1\% of time (8.87 minutes on average per session) on manual interactions, resulting in an average of 28.6 manual interactions per session. Additionally, every participant performed more than 70\% of their interactions manually. These results indicate that despite receiving diverse guidance, participants maintained a strong preference for manual control and agency. Besides, we observe that chatting with AI and  in‑situ suggestions were also commonly used: they appeared in 86.2\% and 82.8\% of sessions, respectively, and accounted for 61.5\% and 35.2\% of guidance-application events. Among participants who used these features, the average usage was 4.5 chat messages and 2.7 in‑situ suggestion applications per session. By contrast, peripheral suggestions were rarely applied: they were used in only 13.8\% of sessions and accounted for just 3.3\% of guidance-application events. Furthermore, we have noticed that the top 2 suggestion types were used together in 58.6\% sessions, and 10.3\% sessions used all the suggestions types. Interestingly, there are 5 sessions (17.2\%) in which only in-situ suggestions were the only guidance used, and there are 3 sessions where only chatting with AI were used, revealing diversity in participants' preference of the suggestion types.

\textbf{Usage patterns.}
Several usage patterns emerged from the detailed participant trials. One of the most common is illustrated in \autoref{fig:p15a}, demonstrated here with an example workflow from Participant 15 on Task 1. The user began with data extraction through an iterative collaboration with the AI: (1) building data examples manually, (2) receiving automatic table completions via in-situ suggestions from the AI, and (3) accepting those suggestions. Next, the user proceeded to data wrangling, predominantly using chat interactions, peripheral guidance, or manual adjustments. During this phase, in-situ suggestions were infrequently generated or accepted. Finally, participants created data visualizations before making their final decisions, often revisiting the wrangling stage upon noticing issues. Throughout this stage, most participants performed the majority of the work manually. Such a pattern revealed distinct behavior of participants between different stages and tasks.
We further asked participants to reason about their interaction choices in the post-study interview. As P3 noted, he started from an example to ``\textit{provide the AI with a blueprint of what should be done to avoid redundant extraction}''. He  also manually created visualizations because it ``\textit{mostly depended on personal understanding}''. P9 added that the task was ``\textit{intuitive and easily done within a few drags}''. By contrast, most participants explained that they later turned to use chatting for data wrangling to ``\textit{save effort}'' (P14) or because they were not sure if the subtask could be done manually.

\begin{figure*}[t]
  \centering
\includegraphics[width=0.9\linewidth]{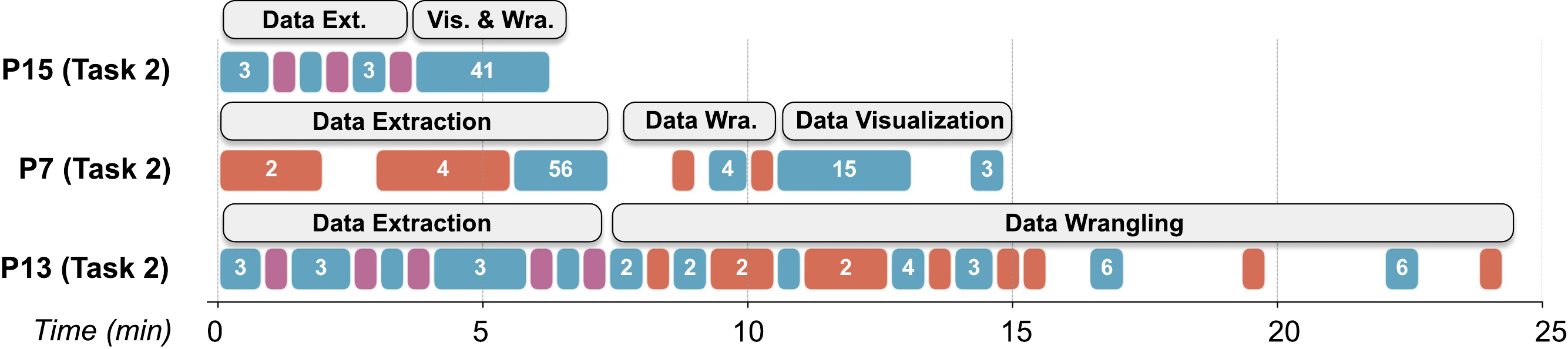}
  \caption{A visualization of P15, P7, and P13's workflows on Task 2.}
  \Description{In P15's workflow on Task 2, only direct manipulation and in-situ guidance were used. In P7's workflow on Task 2, only direct manipulation and chatting with AI were used. In P13's workflow on Task 2, only two stages existed: Data Extraction and Data Wrangling.}
  \label{fig:p15713}
\end{figure*}

We have also observed a few patterns that deviate from the common ones. For instance, as shown in \autoref{fig:p15713}, P15 manually did all subtasks in data visualization and wrangling without turning to AI in Task 2. P7, however, followed an opposite approach where he merely wrote lengthy prompts that included multiple subtasks instead of further decomposing the task and refining data instances gradually in Task 2. Moreover, not all participants followed the three stages mentioned above. For example, P13 reached his final decision by chatting with AI to calculate a score for each item based on a custom formula in Task 2, without making any visualizations.

In addition, we have noticed a few interesting micro patterns in different subtasks within the phases. For instance, while most users started by extracting all items from the source webpage, there are a few users who first sorted or filtered items using native interactions on the web (e.g., the product filtering panel on the left of Amazon) before creating the very first table (P5, P7, and P9 in Task 2). P9 even explained in the post-study interview that she felt the native interactions natural and hoped that WebSeek's interactions interlinked with the native ones more closely. During the data extraction, most participants deliberately stopped to wait for the generation in-situ suggestions, a phenomenon that relates with prior studies\cite{sperrle2024wizard}. By contrast, P1, P2, and P7, proactively wrote prompts to ask for table completion. This strategic diversity was most evident in the table joining subtask (e.g., joining the two tables in Task 1 or integrating product  data from different platforms  in Task 2): 
of the sessions that involved a join, 6 used chat, 4 used direct manipulation, and only 3 used the proactive peripheral suggestion, while another 3 manually integrated data without an explicit join operation. This variety underscores the importance of providing multiple interaction modalities to support a wide range of user preferences and problem-solving styles.


\subsubsection{A Symbiotic Workflow: Blending Manual Control with Multi-Faceted AI Assistance}
Participants unanimously leveraged a combination of direct manipulation and AI assistance to complete their tasks. The vast majority of participants found direct manipulation and in-situ suggestions to be the most useful and essential components of their workflow.

\textbf{Direct manipulation} was considered indispensable. Participants valued the sense of control it provided, with one stating it was essential for ``\textit{small fixes and adjustments where waiting for the AI would be too slow}'' (P9). 
P3 similarly stated, ``\textit{A system like this must have manual features. They made me feel like the human has full control in all pivot steps within the task while AI serves as an assistant.}''
This control was closely linked to trust and accuracy. As P14 explained, ``\textit{I prefer to do some steps myself so the system can understand my intent and complete it for me. The accuracy is much higher this way. If I just use my words to describe what I want, I'm not sure it can get my meaning.}'' Several participants felt they had a better understanding of the system's manual features than the ``\textit{capability boundaries of the LLM}'' (P7), making manual actions feel more reliable and predictable. This was especially true for small-scale tasks (P11, P15) or when users wanted to provide examples to the AI, a strategy one user found effective by building example data instances: ``\textit{I believe that the workflow is natural as I usually give LLMs examples when using them}'' (P6).

\textbf{In-situ suggestions} were also universally praised for their ability to reduce tedious, repetitive work by all participants, particularly the efficiency it brought to web data extraction. Most users found these suggestions accurate and intuitive, giving them a sense of ``\textit{what you see is what you get}'' (P13). Several participantss also valued that the suggestions could complete multiple subtasks at once, not just simple data scraping. For example, P3 said, ``\textit{In the second task I actually forgot to name the columns in the table. If I had hit save, everything later would have been a mess, but the completion feature filled in the column names for me, so I found it very useful and the overall results were accurate.}''
However, feedback also highlighted a desire for more intelligent control over the granularity of these completions. For instance, P5 wished for a ``\textit{slider to manually control the overhead of the suggestions}'', while others (P7, P10) distinguished between suggestions that served their ``\textit{local intent}'' (completing a small part of a table) versus their ``\textit{global intent}'' (advancing the entire task), urging that suggestions of different categories be put in different places. 

In contrast, \textbf{peripheral AI suggestions} were used more selectively. Participants who were unsure of what to do next or ``\textit{needed inspiration}'' (P8, P11) found them helpful as a guide. However, many who had a clear plan found them unnecessary (P7, P11, P12, P13) and cited the high cognitive cost of checking the suggestions one by one or shifting their attention to the suggestions panel (P9, P14). As P9 noted, ``\textit{I applied a suggestion mostly when I was about to chat with the LLM and I noticed it. In other time I would never check it when I was devoted to working on the view below.}'' She further suggested that the peripheral suggestions be put as close to the workspace as possible. ``\textit{For example, a bubble can be added to the side of the workspace, notifying me when new suggestion are available. In this way, it can attract my attention and make me more willing to inspect the suggestions.}''
Some participants (P4, P8) also criticized that the suggestions were trivial or not aligned with their intents. As P10 put it, ``\textit{there is no need to recommend sortings or filterings as they were manageable through only a few clicks}''. Some users were also skeptical of the AI's ability to understand their personal priorities (P5). For instance, he believed the \textit{price} of the products to be the most crucial criterium when making a visualization and said that the AI suggestions would ``\textit{never get this point}''. Interestingly, we also observed a few participants who accepted the suggestions ``\textit{cognitively}'' rather than ``\textit{physically}'', using the suggestion as a reference and performing the same or a similar  operation to the suggested one manually rather than directly clicking on the \textit{apply} button. As P3 noted, ``\textit{While they did not provide direct insights, they inspired me to think about whether I should do like that, even if the suggested operations were not aligned with the study tasks.}''
P12 also explained, ``\textit{I saw its suggestion, but I subconsciously did it manually anyway.}''

Finally, \textbf{chatting with AI} was predominantly used as a tool for complex operations that were difficult or impossible to perform manually, such as semantic substring extraction (P6, P9), or as a ``last resort'' when manual operations failed (P8, P11). While almost all participants agreed on the intuitiveness and efficiency of chatting, some of them shared their complaints with us. P2, for instance, desired an end-to-end style of data analysis where data instances and analysis results were generated as a whole directly from the prompt. In addition, echoing feedback on in-situ suggestions, participants asked for better control over the granularity and overhead of the LLM’s work. As P13 noted: ``\textit{When I said ‘add a column’, it just added an empty column without filling it with data which could have been easily inferred}''.

In post-study interviews, we asked participants to identify when they preferred direct manipulation versus conversational AI interaction. We observed that participants favored direct manipulation when tasks were perceived as simple (e.g., editing a single cell), when accuracy was critical (e.g., filtering), or when correcting AI errors. Direct manipulation was also preferred during thoughtful exploration, where manual control aided thinking. Conversely, participants preferred chatting when manual execution felt laborious or inefficient, or when they were unsure if they could perform the operation manually. For instance, participants generally utilized chat for complex tasks like normalizing inconsistent entities or joining tables. However, they often reverted to manual editing during the initial stages of a task while their goals were still ambiguous. Notably, P3 remarked that these preferences were dynamic, shifting as he gained a better understanding of the AI's capabilities. Overall, our findings suggest that participants sought to balance automation with control, leveraging AI to bypass tedious execution while retaining manual oversight for tasks demanding accuracy and nuanced decision-making.

\subsubsection{The Power of a Unified, Data-Centric Workspace}
A major theme that emerged when comparing WebSeek to both traditional workflows and other AI agents was the significant benefit of its integrated environment, which eliminated the need for context and platform switching. P14 articulated this clearly: ``\textit{[Traditionally], I have to constantly switch between various tools to check data before I can take the next step. But on this platform, all the operations are integrated.}'' P3 similarly strengthened that ``\textit{The fact that it is a browser extension is everything}''. Meanwhile, many participants mentioned that they had a strong grasp of their data within WebSeek. P7 highlighted that ``\textit{the data always stays in my workspace, providing context for the agent,}'' and appreciated the ability to ``\textit{precisely roll back to a previous step if I want to explore another possibility, which you can't do with an agent that doesn't have this data persistence.}'' This sentiment was echoed by others, who noted that WebSeek's process was ``\textit{clearer, with every step controllable, checkable, and modifiable}'' (P12). 
The ability to see the ``\textit{full picture of the data while chatting}'' was described as more comfortable and powerful than a standard, isolated chatbot interface (P2, P6).

Additionally, we have received mixed feedback on the design of freely arranging data instances on a canvas interface. P13 noted that the canvas enabled comparing visualizations in parallel, while P11 preferred a classical file system-like UI to avoid visual clusters caused by messy instances. P14 suggested that the relations between data instances  (e.g., derivations and computations) be visualized graphically to facilitate understanding data provenance and organizing thoughts. Besides, P1 hoped that instances of diverse modalities (e.g., web screenshots) could be supported for better expressiveness and flexibility.

\subsubsection{Sense of Confidence Through Transparency and Control}
The combination of direct manipulation and a persistent data canvas directly translated to user confidence. Participants consistently linked their confidence to the feeling of control and the ability to personally verify the system's output. As P6 stated, the ``\textit{manual process gives a stronger sense of control and confidence}.'' This was reinforced by P14, who felt confident ``\textit{when I can observe with my own eyes that the intermediate results are correct.}'' Transparency into the AI's operations was another key factor affecting the user confidence. P9 appreciated the ability to look into the detailed workflow of peripheral AI guidance and indicated that it contributed to her confidence. Conversely, participants worried that a "black box" agent like the LLM for generating in-situ suggestions  might not have extracted all the data completely (P14). One of the most insightful comments came from P7, who noted that compared to asking AI for recommendations,  the process of working with data and building visualizations within WebSeek gave him ``\textit{a stronger cognitive understanding of the data}'' that strengthened his confidence on the final decision. To further improve this, P7 suggested that even in the chat, the system should explicitly show the tool-calling process to make the AI's actions fully transparent.

\section{Discussion}
\subsection{Rethinking Web Agents: From Conversational Delegation to Data-Centric Collaboration}
Our findings contribute to a broader conversation about the future of agentic web automation. The dominant paradigm, exemplified by systems like ChatGPT Agent, is one of conversational delegation: the user delegates a high-level goal, and the AI executes it, often as a black box. While this approach offers a promising path towards efficiency, its inherent structure presents challenges for complex, data-driven decision making where nuance, trust, and iterative refinement are paramount. As observed in our study, participants highly valued transparency, direct control over their data, and the ability to verify each step—qualities often at odds with a purely delegative model. We hence argue that for tasks where data plays a pivotal role, the interaction model itself must be centered on data. Instead of hiding data within an autonomous process, a data-centric model makes it the primary unit of interaction—a tangible, inspectable, and reusable artifact that the user and AI collaborate on. Our study shows that such data instances can serve as a  \textit{mutual context} between users and AI, reducing the effort of data-related communication.

Moreover, we suggest that in this new paradigm, users should be granted full control through a suite of interaction modalities, from direct manipulation for fine-grained control to natural language for high-level commands. Within this partnership, the AI's role, whether proactive or reactive, must strictly adhere to a ``User-Centered AI'' philosophy~\cite{narechania2025agentic},
i.e., to assist and not to lead. We envision that the user must always be the one to determine the critical pivot points in the decision making process, while the AI should strive to meet the diverse data needs with different degrees of granularity and personalization as demonstrated by our framework and study.

\subsection{Infering Guidance Without an Evident Cue}
Our framework for proactive assistance is intentionally cue-driven, designed to trigger suggestions based on concrete, observable user actions. For example, the system suggests table completion only after the user has provided a few manual examples. This approach prioritizes suggestion accuracy and respects user agency, ensuring that most proactive interventions are highly relevant to the user's immediate, or     ``local'' intent. Our study results confirm that users found this type of assistance highly effective and intuitive.
However, this reliance on evident cues creates an inherent limitation: the system may fail to provide valuable guidance when no explicit cue is available. This presents a ``cold-start'' problem for proactivity, particularly for suggestions that align with a user's high-level ``global intent'' rather than their immediate actions. For instance, a user manually filling the initial rows of a table might be focused on the local task of data extraction, never providing a clear cue to trigger a more globally useful suggestion like ``\textit{this dataset could be joined with product reviews from another source to enrich your analysis}''.
Furthermore, this cue-driven model struggles to offer guidance on problems that users themselves may not have noticed. A critical part of data analysis is identifying and correcting hidden data quality issues. Our current system is unlikely to suggest, for example, ``\textit{this column of product weights contains mixed units (lbs and kg) and should be normalized}'', because the user's actions might not provide any signal that they are aware of or concerned with this problem. Overcoming this limitation would require the AI to perform more autonomous, background analysis of the data artifacts to identify latent opportunities for improvement. While this introduces the risk of generating less relevant suggestions, it points to a crucial direction for future work: developing techniques to balance highly-relevant, cue-driven local assistance with more speculative but potentially transformative global guidance.

\subsection{Bridging the Gap Between a Data-Driven web extension and Native Web Interactions}
Built as a data-instance-oriented browser extension, WebSeek centers most of its interaction design around data for a more transparent, controllable, and smooth experience of data-driven decision making on the web. Our study demonstrated that by providing a unified space for data, users felt confident and in control, benefiting from a seamless, human-in-the-loop analysis experience. However, in abstracting data away from its source, we admittedly paid less attention to integrating with the rich interactive affordances of the native web environment itself.
Participants in our study often began their tasks by using familiar, native web interactions before extracting any data into WebSeek, such as the filtering panel on an e-commerce site. As P9 insightfully noted, these interactions felt natural, leading her to wish that ``\textit{WebSeek's interactions were more closely interlinked with the native ones.}'' Likewise, other participants hoped that the system could automatically navigate into the the detail page or integrate clean, analysis-ready data from other web sources without even opening them. This highlights a fundamental design tension: how can a system provide a powerful, external analysis environment without creating a disruptive seam between it and the original web content? Future work should explore how to better balance these external and native interactions, making tools like WebSeek more environmentally aware and malleable~\cite{malleable}, adapting to and augmenting the user's existing web workflows rather than completely replacing them.

Furthermore, we identify a key opportunity to improve transparency by strengthening the visual link between data artifacts and their web sources. While WebSeek currently supports navigating back to a data item's origin, this link is not persistent or always visible. As noted by one experienced data analyst, P3, this ambiguity can erode confidence, especially after complex data transformations have occurred. A powerful solution, inspired by P3's feedback, would be to create a persistent visual overlay on the source webpage. He envisioned an ``\textit{overview on the webpage showing which elements have been extracted and where they are in the system}''. This explicit mapping would provide robust data provenance, allowing users to instantly verify their data and trace its lineage, even after it has been cleaned, merged, and transformed, thus tightening the feedback loop between the web and the canvas.

\subsection{Data as Tangible, Malleable Instances on a Canvas}
Contrary to the traditional strategy of treating data instances as concrete items like files, WebSeek makes data instances tangible to encourage the construction of incomplete or temporary data examples and comparing instances. 
It suggests a move away from abstract commands towards the direct creation and manipulation of tangible data instances on an interactive canvas. This shift towards tangible, constructive interfaces for data is gaining traction in the HCI community. For example, Table Illustrator~\cite{huang2024table} reimagines table authoring by treating layout components as tangible ``puzzles'' that users can combine and configure.  Likewise, DeckFlow~\cite{croisdale2025deckflow} allows users to manipulate text, image, and audio cards on an infinite canvas for multi-modal content generation, facilitating task decomposition and open-ended generation. This paradigm is also applicable to code files for software engineering as demonstrated in Code Canvas~\cite{deline2010code}. WebSeek applies a similar philosophy and shows that it can be beneficial to the general data analysis or decision making process. First, it lowers the cognitive load by making the data and its state persistently visible, allowing users to ``think'' by arranging and structuring artifacts on the canvas. Second, it encourages exploration; users can freely duplicate, branch, and transform data instances without fear of losing their work.
Finally, it opens the door for richer, multi-modal interactions. As the saying goes, ``\textit{everything is data}''~\cite{yunita2022everything},  these data instances can be made increasingly malleable. We envision a future where these tangible artifacts can be embedded within one another—for instance, embedding a table into a sketch for editing, or dropping a web screenshot into the canvas for direct table creation—to create a rich, hierarchical, and deeply personal space for sensemaking.


\subsection{Limitations and Future Work}
Our work has several limitations that point to promising directions for future research.

First, our current technical approach for providing context to the LLM is not optimized for scale. To generate relevant suggestions, WebSeek includes the raw HTML of the current webpage and the state of all data instances on the canvas in its prompt. This approach, while effective for our study's tasks, can be computationally expensive and inefficient, potentially leading to high latency and cost with very large or complex webpages. Future work should explore more sophisticated context management techniques, such as using vector embeddings for semantic search, dynamic context pruning, or developing smaller, fine-tuned models for specific sub-tasks like data extraction. 

Second, while the canvas-based interface was effective for the tasks in our study, its usability may degrade as the number of data artifacts grows significantly. Future work could explore more advanced organizational paradigms for the canvas to manage complexity at scale, such as hierarchical grouping, portals between different views, or relationship graphs that visually link related instances, as suggested by some of our participants. 


Finally, some threats exist to the validity of our studies. For the technical study, the benchmark size was limited due to the labor-intensive nature of our human-in-the-loop procedure. In addition, although we manually infused data quality issues to mimic real-world dirty data, these synthetic pages may differ from live websites in terms of DOM complexity, lacking elements such as extremely nested elements, aggressive advertising, or complex dynamic content loading. In our future work, we plan to conduct a study with a larger scale in the wild. For the user study, the tasks in our study merely cover a limited scope in the data analysis landscape and may not accurately reflect user behavior in more rigorous or large-scale tasks, such as benchmark generation and scientific data analysis. Besides, the participant response biases~\cite{pr-bias} could not be fully eliminated. Additionally, all participants were students, and the size of the corpus was limited, which may bias the findings. Lastly, a more comprehensive evaluation of our proactive AI guidance framework (Section \ref{proactiveassistance}) out of WebSeek could have been conducted for better assessment of its generalizability. We leave this study as future work.

\section{Conclusion}
We presented WebSeek, a mixed-initiative browser extension that reimagines human-AI collaboration for web-based, data-driven tasks. Moving beyond the limitations of purely conversational agents, WebSeek introduces a data-first paradigm where users directly manipulate tangible data artifacts on an interactive canvas. This process is augmented by a flexible LLM-powered assistant, guided by a principled design space for proactive and reactive guidance. A user study demonstrated the effectiveness and usability of WebSeek, showing that it enables users to successfully complete complex tasks while reporting high levels of confidence and sense of agency and control. Ultimately, our work argues for a shift in designing web agents, i.e., from opaque, conversational delegation to transparent, data-centric collaboration, opening a rich design space for future human-AI data curation, analysis, or decision making tools on the web.

\bibliographystyle{ACM-Reference-Format}
\bibliography{reference}

@String{Computing = "Computing" }

@String{Computer = "{IEEE} Computer" }

@String{Springer = "Springer-Verlag" }

@Misc{ChatGPT-Agent,
  author={OpenAI}, 
  year={2025}, 
  howpublished = {\url{https://openai.com/index/introducing-chatgpt-agent/}},
  note         = {Accessed: August 19, 2025},
  title        = {{ChatGPT} {Agent}.},
}

@Misc{GenSpark,
  author={GenSpark}, 
  year={2025}, 
  howpublished = {\url{https://www.genspark.ai/}},
  note         = {Accessed: August 19, 2025},
  title        = {{GenSpark} {Super} {Agent}.},
}

@Misc{Harpa-ai,
  author={Harpa.ai}, 
  year={2025}, 
  howpublished = {\url{https://harpa.ai/}},
  note         = {Accessed: August 19, 2025},
  title        = {{Harpa.AI}.},
}

@Misc{Merlin-ai,
  author={Foyer Tech Inc.}, 
  year={2025}, 
  howpublished = {\url{https://www.getmerlin.in/}},
  note         = {Accessed: August 19, 2025},
  title        = {{Merlin.AI}.},
}

@Misc{Skyvern,
  author={Ikonomos Inc.}, 
  year={2025}, 
  howpublished = {\url{https://www.skyvern.com/}},
  note         = {Accessed: August 19, 2025},
  title        = {Skyvern.},
}

@Misc{Manus-ai,
  author={Manus AI}, 
  year={2025}, 
  howpublished = {\url{https://manus.im/}},
  note         = {Accessed: August 19, 2025},
  title        = {{Manus.AI}.},
}

@Misc{AgenticSeek,
  author={Fosowl}, 
  year={2025}, 
  howpublished = {\url{https://github.com/Fosowl/agenticSeek}},
  note         = {Accessed: August 19, 2025},
  title        = {{AgenticSeek: Private, Local Manus Alternative.}.},
}

@Misc{GenSpark-Browser,
  author={GenSpark}, 
  year={2025}, 
  howpublished = {\url{https://www.genspark.ai/browser}},
  note         = {Accessed: August 19, 2025},
  title        = {{GenSpark Browser}.},
}

@Misc{Copilot,
  author={GitHub}, 
  year={2025}, 
  howpublished = {\url{https://github.com/features/copilot}},
  note         = {Accessed: August 19, 2025},
  title        = {{GitHub Copilot}.},
}

@Misc{AnythingLLM,
  author={Mintplex Labs}, 
  year={2025}, 
  howpublished = {\url{https://github.com/Mintplex-Labs/anything-llm}},
  note         = {Accessed: August 19, 2025},
  title        = {{AnythingLLM}.},
}

@inproceedings{he2024webvoyager,
    title = "{W}eb{V}oyager: Building an End-to-End Web Agent with Large Multimodal Models",
    author = "He, Hongliang  and
      Yao, Wenlin  and
      Ma, Kaixin  and
      Yu, Wenhao  and
      Dai, Yong  and
      Zhang, Hongming  and
      Lan, Zhenzhong  and
      Yu, Dong",
    booktitle = "Proceedings of the Annual Meeting of the Association for Computational Linguistics",
    month = aug,
    year = "2024",
    url = "https://aclanthology.org/2024.acl-long.371/",
    doi = "10.18653/v1/2024.acl-long.371",
    pages = "6864--6890",
}

@article{agentic-web,
  title={Agentic Web: Weaving the Next Web with AI Agents},
  author={Yang, Yingxuan and Ma, Mulei and Huang, Yuxuan and Chai, Huacan and Gong, Chenyu and Geng, Haoran and Zhou, Yuanjian and Wen, Ying and Fang, Meng and Chen, Muhao and others},
  journal={arXiv preprint arXiv:2507.21206},
  year={2025}
}

@article{song2024mmac,
  title={{MMAC-Copilot}: Multi-modal Agent Collaboration Operating Copilot},
  author={Song, Zirui and Li, Yaohang and Fang, Meng and Li, Yanda and Chen, Zhenhao and Shi, Zecheng and Huang, Yuan and Chen, Xiuying and Chen, Ling},
  journal={arXiv preprint arXiv:2404.18074},
  year={2024}
}

@article{shao2025future,
  title={Future of Work with AI Agents: Auditing Automation and Augmentation Potential across the US Workforce},
  author={Shao, Yijia and Zope, Humishka and Jiang, Yucheng and Pei, Jiaxin and Nguyen, David and Brynjolfsson, Erik and Yang, Diyi},
  journal={arXiv preprint arXiv:2506.06576},
  year={2025}
}

@book{cronin2024understanding,
  title={Understanding Generative AI Business Applications},
  author={Cronin, Irena},
  year={2024},
  publisher={Springer}
}

@article{huang2023interactive,
  title={Interactive Table Synthesis with Natural Language},
  author={Huang, Yanwei and Zhou, Yunfan and Chen, Ran and Pan, Changhao and Shu, Xinhuan and Weng, Di and Wu, Yingcai},
  journal={IEEE Transactions on Visualization and Computer Graphics},
  volume={30},
  number={9},
  pages={6130--6145},
  year={2024},
  publisher={IEEE},
  doi={10.1109/TVCG.2023.3329120}
}

@article{2019-agency-automation,
  title = {Agency plus Automation: Designing Artificial Intelligence into Interactive Systems},
  author = {Heer, Jeffrey},
  journal = {Proceedings of the National Academy of Sciences},
  year = {2019},
  volume = {116},
  number = {6},
  pages = {1844--1850},
  url = {https://idl.uw.edu/papers/agency-automation},
  doi = {10.1073/pnas.1807184115}
}

@inproceedings{paden2023biasbuzz,
author = {Paden, Jamal R and Narechania, Arpit and Endert, Alex},
title = {{BiasBuzz: Combining Visual Guidance with Haptic Feedback to Increase Awareness of Analytic Behavior during Visual Data Analysis}},
year = {2024},
url = {https://doi.org/10.1145/3613905.3651064},
doi = {10.1145/3613905.3651064},
booktitle = {Extended Abstracts of the CHI Conference on Human Factors in Computing Systems},
}

@inproceedings{srinivasan2021collecting,
  title={Collecting and Characterizing Natural Language Utterances for Specifying Data Visualizations},
  author={Srinivasan, Arjun and Nyapathy, Nikhila and Lee, Bongshin and Drucker, Steven M and Stasko, John},
  booktitle={Proceedings of the CHI Conference on Human Factors in Computing Systems},
  pages={1--10},
  year={2021},
  doi = {10.1145/3411764.3445400},
}

@inproceedings{masson2024directgpt,
  title={{DirectGPT}: A Direct Manipulation Interface to Interact with Large Language Models},
  author={Masson, Damien and Malacria, Sylvain and Casiez, G{\'e}ry and Vogel, Daniel},
  booktitle={Proceedings of the CHI Conference on Human Factors in Computing Systems},
  pages={1--16},
  year={2024},
  doi = {10.1145/3613904.3642462},
}

@article{ma2023laser,
  title={{LASER}: LLM Agent with State-Space Exploration for Web Navigation},
  author={Ma, Kaixin and Zhang, Hongming and Wang, Hongwei and Pan, Xiaoman and Yu, Wenhao and Yu, Dong},
  journal={NeurIPS},
  year={2023}
}

@inproceedings{zhang2023you,
    title = "You Only Look at Screens: Multimodal Chain-of-Action Agents",
    author = "Zhang, Zhuosheng  and
      Zhang, Aston",
    editor = "Ku, Lun-Wei  and
      Martins, Andre  and
      Srikumar, Vivek",
    booktitle = "Findings of the Association for Computational Linguistics",
    month = aug,
    year = "2024",
    url = "https://aclanthology.org/2024.findings-acl.186/",
    doi = "10.18653/v1/2024.findings-acl.186",
    pages = "3132--3149",
}

@inproceedings{webagent-survey,
  title={A Survey of {WebAgents}: Towards Next-Generation AI Agents for Web Automation with Large Foundation Models},
  author={Ning, Liangbo and Liang, Ziran and Jiang, Zhuohang and Qu, Haohao and Ding, Yujuan and Fan, Wenqi and Wei, Xiao-yong and Lin, Shanru and Liu, Hui and Yu, Philip S and others},
  booktitle={Proceedings of the 31st ACM SIGKDD Conference on Knowledge Discovery and Data Mining},
  pages={6140--6150},
  year={2025}
}

@article{doan2009information,
  title={Information Extraction Challenges in Managing Unstructured Data},
  author={Doan, AnHai and Naughton, Jeffrey F and Ramakrishnan, Raghu and Baid, Akanksha and Chai, Xiaoyong and Chen, Fei and Chen, Ting and Chu, Eric and DeRose, Pedro and Gao, Byron and others},
  journal={ACM SIGMOD Record},
  volume={37},
  number={4},
  pages={14--20},
  year={2009},
}

@Misc{beautiful-soup,
year={2025}, 
  howpublished = {\url{https://www.crummy.com/software/BeautifulSoup/}},
  note         = {Accessed: August 20, 2025},
  title        = {Beautiful Soup Library. },
}

@article{dalvi2011automatic,
author = {Dalvi, Nilesh and Kumar, Ravi and Soliman, Mohamed},
title = {Automatic wrappers for large scale web extraction},
year = {2011},
volume = {4},
number = {4},
issn = {2150-8097},
url = {https://doi.org/10.14778/1938545.1938547},
doi = {10.14778/1938545.1938547},
journal = {Proceedings of the VLDB Endowment},
month = jan,
pages = {219–230},
}

@article{ortona2015wadar,
  title={{WADaR}: Joint Wrapper and Data Repair},
  author={Ortona, Stefano and Orsi, Giorgio and Buoncristiano, Marcello and Furche, Tim},
  journal={Proceedings of the VLDB Endowment},
  volume={8},
  number={12},
  pages={1996--1999},
  year={2015},
  publisher={VLDB Endowment}
}

@article{inala2017webrelate,
  title={{WebRelate}: integrating web data with spreadsheets using examples},
  author={Inala, Jeevana Priya and Singh, Rishabh},
  journal={Proceedings of the ACM on Programming Languages},
  volume={2},
  number={POPL},
  pages={1--28},
  year={2017},
}

@inproceedings{le2014flashextract,
  title={{FlashExtract}: a framework for data extraction by examples},
  author={Le, Vu and Gulwani, Sumit},
  booktitle={Proceedings of the 35th ACM SIGPLAN Conference on Programming Language Design and Implementation},
  pages={542--553},
  year={2014}
}

@inproceedings{barman2016ringer,
  title={Ringer: web automation by demonstration},
  author={Barman, Shaon and Chasins, Sarah and Bodik, Rastislav and Gulwani, Sumit},
  booktitle={Proceedings of the 2016 ACM SIGPLAN International Conference on Object-Oriented Programming, Systems, Languages, and Applications},
  pages={748--764},
  year={2016}
}

@inproceedings{chasins2018rousillon,
  title={Rousillon: Scraping distributed hierarchical web data},
  author={Chasins, Sarah E and Mueller, Maria and Bodik, Rastislav},
  booktitle={Proceedings of the 31st Annual ACM Symposium on User Interface Software and Technology},
  pages={963--975},
  year={2018},
  doi = {10.1145/3242587.3242661}
}

@inproceedings{pu2022semanticon,
  title={{SemanticOn}: Specifying Content-Based Semantic Conditions for
Web Automation Programs},
  author={Pu, Kevin and Fu, Rainey and Dong, Rui and Wang, Xinyu and Chen, Yan and Grossman, Tovi},
  booktitle={Proceedings of the 35th Annual ACM Symposium on User Interface Software and Technology},
  pages={1--16},
  year={2022},
  doi = {10.1145/3526113.3545691}
}

@inproceedings{satyanarayan2014lyra,
  title={Lyra: An Interactive Visualization Design Environment},
  author={Satyanarayan, Arvind and Heer, Jeffrey},
  booktitle={Computer graphics forum},
  volume={33},
  number={3},
  pages={351--360},
  year={2014},
  organization={Wiley Online Library}
}

@article{zong2020lyra,
  title={Lyra 2: Designing Interactive Visualizations by Demonstration},
  author={Zong, Jonathan and Barnwal, Dhiraj and Neogy, Rupayan and Satyanarayan, Arvind},
  journal={IEEE Transactions on Visualization and Computer Graphics},
  volume={27},
  number={2},
  pages={304--314},
  year={2020},
  publisher={IEEE},
  doi={10.1109/TVCG.2020.3030367}
}

@inproceedings{profiler,
author = {Kandel, Sean and Parikh, Ravi and Paepcke, Andreas and Hellerstein, Joseph M. and Heer, Jeffrey},
title = {Profiler: Integrated Statistical Analysis and Visualization for Data Quality Assessment},
year = {2012},
publisher = {ACM},
doi = {10.1145/2254556.2254659},
booktitle = {Proceedings of the International Working Conference on Advanced Visual Interfaces},
pages = {547--554},
}

@inproceedings{wrangler,
  author    = {Sean Kandel and
               Andreas Paepcke and
               Joseph M. Hellerstein and
               Jeffrey Heer},
  title     = {Wrangler: Interactive Visual Specification of Data Transformation Scripts},
  year      = {2011},
  booktitle = {Proceedings of the CHI Conference on Human Factors in Computing Systems},
  pages     = {3363--3372},
  
  doi       = {10.1145/1978942.1979444}
}

@Misc{trifacta,
  author       = {{Trifacta, Inc.}},
  howpublished = {\url{https://www.trifacta.com}},
  note         = {Accessed: August 20, 2025},
  title        = {Trifacta.},
}

@Misc{openrefine,
  author       = {{OpenRefine, Inc.}},
  title        = {Open {Refine}.},
  howpublished = {\url{https://openrefine.org}},
  note         = {Accessed: August 20, 2025},
}

@article{satyanarayan2016vega,
  title={Vega-Lite: A Grammar of Interactive Graphics},
  author={Satyanarayan, Arvind and Moritz, Dominik and Wongsuphasawat, Kanit and Heer, Jeffrey},
  journal={IEEE Transactions on Visualization and Computer Graphics},
  volume={23},
  number={1},
  pages={341--350},
  year={2016},
  publisher={IEEE},
  doi={10.1109/TVCG.2016.2599030}
}

@article{wongsuphasawat2015voyager,
  title={Voyager: Exploratory Analysis via Faceted Browsing of Visualization Recommendations},
  author={Wongsuphasawat, Kanit and Moritz, Dominik and Anand, Anushka and Mackinlay, Jock and Howe, Bill and Heer, Jeffrey},
  journal={IEEE Transactions on Visualization and Computer Graphics},
  volume={22},
  number={1},
  pages={649--658},
  year={2015},
  publisher={IEEE},
  doi={10.1109/TVCG.2015.2467191}
}

@inproceedings{wongsuphasawat2017voyager,
  title={Voyager 2: Augmenting Visual Analysis with Partial View Specifications},
  author={Wongsuphasawat, Kanit and Qu, Zening and Moritz, Dominik and Chang, Riley and Ouk, Felix and Anand, Anushka and Mackinlay, Jock and Howe, Bill and Heer, Jeffrey},
  booktitle={Proceedings of the CHI Conference on Human Factors in Computing Systems},
  pages={2648--2659},
  year={2017},
  doi = {10.1145/3025453.3025768}
}

@phdthesis{narechania2024dissertation,
  title={Designing, Developing, and Democratizing Guidance for Visual Analytics},
  author={Narechania, Arpit Ajay},
  year={2024},
  school={Georgia Institute of Technology}
}

@inproceedings{cao2022visguide,
  title={{VisGuide}: User-Oriented Recommendations for Data Event Extraction},
  author={Cao, Yu-Rong and Li, Xiao-Han and Pan, Jia-Yu and Lin, Wen-Chieh},
  booktitle={Proceedings of the CHI Conference on Human Factors in Computing Systems},
  pages={1--13},
  year={2022},
  doi = {10.1145/3491102.3517648},
}

@article{wang2025mobile,
      title={Mobile-Agent-E: Self-Evolving Mobile Assistant for Complex Tasks},
      author={Wang, Zhenhailong and Xu, Haiyang and Wang, Junyang and Zhang, Xi and Yan, Ming and Zhang, Ji and Huang, Fei and Ji, Heng},
      journal={NeurIPS},
      year={2025}
    }

@article{abuelsaad2024agent,
  title={{Agent-E}: From Autonomous Web Navigation to Foundational Design Principles in Agentic Systems},
  author={Abuelsaad, Tamer and Akkil, Deepak and Dey, Prasenjit and Jagmohan, Ashish and Vempaty, Aditya and Kokku, Ravi},
  journal={arXiv preprint arXiv:2407.13032},
  year={2024}
}

@article{zhao2025proactiveva,
  title={{ProactiveVA}: Proactive Visual Analytics with LLM-Based UI Agent},
  author={Zhao, Yuheng and Shu, Xueli and Fan, Liwen and Gao, Lin and Zhang, Yu and Chen, Siming},
  journal={IEEE VIS},
  year={2025}
}

@inproceedings{guo2011proactive,
  title={Proactive Wrangling: Mixed-Initiative End-User Programming of Data Transformation Scripts},
  author={Guo, Philip J and Kandel, Sean and Hellerstein, Joseph M and Heer, Jeffrey},
  booktitle={Proceedings of the 24th Annual ACM Symposium on User Interface Software and Technology},
  pages={65--74},
  year={2011},
  doi = {10.1145/2047196.2047205},
}

@inproceedings{peng2019design,
  title={Design and Evaluation of Service Robot's Proactivity in Decision-Making Support Process},
  author={Peng, Zhenhui and Kwon, Yunhwan and Lu, Jiaan and Wu, Ziming and Ma, Xiaojuan},
  booktitle={Proceedings of the CHI Conference on Human Factors in Computing Systems},
  pages={1--13},
  year={2019},
  doi = {10.1145/3290605.3300328},
}

@inproceedings{zhou2025xavier,
  title={Xavier: Toward Better Coding Assistance in Authoring Tabular Data Wrangling Scripts},
  author={Zhou, Yunfan and Cai, Xiwen and Shi, Qiming and Huang, Yanwei and Li, Haotian and Qu, Huamin and Weng, Di and Wu, Yingcai},
  booktitle={Proceedings of the CHI Conference on Human Factors in Computing Systems},
  pages={1--16},
  year={2025},
  doi = {10.1145/3706598.3714239},
}

@inproceedings{proactive-trust,
author = {Kraus, Matthias and Wagner, Nicolas and Minker, Wolfgang},
title = {Effects of Proactive Dialogue Strategies on Human-Computer Trust},
year = {2020},
url = {https://doi.org/10.1145/3340631.3394840},
doi = {10.1145/3340631.3394840},
booktitle = {Proceedings of the 28th ACM Conference on User Modeling, Adaptation and Personalization},
pages = {107–116},
}

@inproceedings{need-help,
author = {Chen, Valerie and Zhu, Alan and Zhao, Sebastian and Mozannar, Hussein and Sontag, David and Talwalkar, Ameet},
title = {Need Help? Designing Proactive AI Assistants for Programming},
year = {2025},
url = {https://doi.org/10.1145/3706598.3714002},
doi = {10.1145/3706598.3714002},
booktitle = {Proceedings of the CHI Conference on Human Factors in Computing Systems},
}

@article{meck2024may,
  title={How May I Interrupt? Linguistic-Driven Design Guidelines for Proactive in-Car Voice Assistants},
  author={Meck, Anna-Maria and Draxler, Christoph and Vogt, Thurid},
  journal={International Journal of Human--Computer Interaction},
  volume={40},
  number={22},
  pages={7517--7531},
  year={2024},
  publisher={Taylor \& Francis}
}

@inproceedings{narechania2023datapilot,
  title={{DataPilot}: Utilizing Quality and Usage Information for Subset Selection during Visual Data Preparation},
  author={Narechania, Arpit and Du, Fan and Sinha, Atanu R and Rossi, Ryan and Hoffswell, Jane and Guo, Shunan and Koh, Eunyee and Navathe, Shamkant B and Endert, Alex},
  booktitle={Proceedings of the CHI Conference on Human Factors in Computing Systems},
  pages={1--18},
  year={2023},
  doi = {10.1145/3544548.3581509},
}

@article{cui2019datasite,
  title={{DataSite}: Proactive visual data exploration with computation of insight-based recommendations},
  author={Cui, Zhe and Badam, Sriram Karthik and Yal{\c{c}}in, M Adil and Elmqvist, Niklas},
  journal={Information Visualization},
  volume={18},
  number={2},
  pages={251--267},
  year={2019},
  publisher={SAGE Publications Sage UK: London, England}
}

@ARTICLE{learning-vis-recommendation,
  author={Li, Yixuan and Qi, Yusheng and Shi, Yang and Chen, Qing and Cao, Nan and Chen, Siming},
  journal={IEEE Transactions on Visualization and Computer Graphics}, 
  title={Diverse Interaction Recommendation for Public Users Exploring Multi-view Visualization using Deep Learning}, 
  year={2023},
  volume={29},
  number={1},
  pages={95-105},
  doi={10.1109/TVCG.2022.3209461}}

@inproceedings{pu2025assistance,
  title={Assistance or Disruption? Exploring and Evaluating the Design and Trade-offs of Proactive AI Programming Support},
  author={Pu, Kevin and Lazaro, Daniel and Arawjo, Ian and Xia, Haijun and Xiao, Ziang and Grossman, Tovi and Chen, Yan},
  booktitle={Proceedings of the CHI Conference on Human Factors in Computing Systems},
  pages={1--21},
  year={2025},
  doi = {10.1145/3706598.3713357},
}

@inproceedings{what-can-you-do,
author = {Liao, Q. Vera and Davis, Matthew and Geyer, Werner and Muller, Michael and Shami, N. Sadat},
title = {What Can You Do? Studying Social-Agent Orientation and Agent Proactive Interactions with an Agent for Employees},
year = {2016},
url = {https://doi.org/10.1145/2901790.2901842},
doi = {10.1145/2901790.2901842},
booktitle = {Proceedings of the 2016 ACM Conference on Designing Interactive Systems},
pages = {264–275},
}

@inproceedings{viswanath2025insights,
  title={Insights for Proactive Agents: Design Considerations, Challenges, and Recommendations},
  author={Viswanath, Anargh and Buschmeier, Hendrik},
  booktitle={Joint Proceedings of the ACM IUI 2025 Workshops},
  year={2025}
}

@inproceedings{pu2025promemassist,
author = {Pu, Kevin and Zhang, Ting and Sendhilnathan, Naveen and Freitag, Sebastian and Sodhi, Raj and Jonker, Tanya R.},
title = {{ProMemAssist}: Exploring Timely Proactive Assistance Through Working Memory Modeling in Multi-Modal Wearable Devices},
year = {2025},
url = {https://doi.org/10.1145/3746059.3747770},
doi = {10.1145/3746059.3747770},
booktitle = {Proceedings of the 38th Annual ACM Symposium on User Interface Software and Technology},
}

@article{narechania2025agentic,
  title={Agentic Enterprise: AI-Centric User to User-Centric AI},
  author={Narechania, Arpit and Endert, Alex and Sinha, Atanu R},
  journal={arXiv preprint arXiv:2506.22893},
  year={2025}
}

@inproceedings{malleable,
author = {Min, Bryan and Chen, Allen and Cao, Yining and Xia, Haijun},
title = {Malleable Overview-Detail Interfaces},
year = {2025},
isbn = {9798400713941},
url = {https://doi.org/10.1145/3706598.3714164},
doi = {10.1145/3706598.3714164},
booktitle = {Proceedings of the CHI Conference on Human Factors in Computing Systems},
}

@inproceedings{muller2019data,
  title={How Data Science Workers Work with Data: Discovery, Capture, Curation, Design, Creation},
  author={Muller, Michael and Lange, Ingrid and Wang, Dakuo and Piorkowski, David and Tsay, Jason and Liao, Q Vera and Dugan, Casey and Erickson, Thomas},
  booktitle={Proceedings of the CHI Conference on Human Factors in Computing Systems},
  pages={1--15},
  year={2019},
  doi = {10.1145/3290605.3300356},
}

@article{kandogan2014data,
  title={From Data to Insight: Work Practices of Analysts in the Enterprise},
  author={Kandogan, Eser and Balakrishnan, Aruna and Haber, Eben M and Pierce, Jeffrey S},
  journal={IEEE Computer Graphics and Applications},
  volume={34},
  number={5},
  pages={42--50},
  year={2014},
  publisher={IEEE}
}

@article{kandel2012enterprise,
  title={Enterprise Data Analysis and Visualization: An Interview Study},
  author={Kandel, Sean and Paepcke, Andreas and Hellerstein, Joseph M and Heer, Jeffrey},
  journal={IEEE Transactions on Visualization and Computer Graphics},
  volume={18},
  number={12},
  pages={2917--2926},
  year={2012},
  publisher={IEEE},
  doi={10.1109/TVCG.2012.219}
}

@ARTICLE{Futz-mosey,
  author={Alspaugh, Sara and Zokaei, Nava and Liu, Andrea and Jin, Cindy and Hearst, Marti A.},
  journal={IEEE Transactions on Visualization and Computer Graphics}, 
  title={Futzing and Moseying: Interviews with Professional Data Analysts on Exploration Practices}, 
  year={2019},
  volume={25},
  number={1},
  pages={22-31},
  doi={10.1109/TVCG.2018.2865040}}

@inproceedings{horvitz1999principles,
  title={Principles of Mixed-Initiative User Interfaces},
  author={Horvitz, Eric},
  booktitle={Proceedings of the CHI Conference on Human Factors in Computing Systems},
  pages={159--166},
  year={1999},
  doi = {10.1145/302979.303030},
}

@article{mackenzie2018fitts,
  title={Fitts’ Law},
  author={MacKenzie, I. Scott},
  journal={The Wiley Handbook of Human Computer Interaction},
  volume={1},
  pages={347--370},
  year={2018},
  publisher={Wiley Online Library}
}

@article{laender2002brief,
  title={A Brief Survey of Web Data Extraction Tools},
  author={Laender, Alberto HF and Ribeiro-Neto, Berthier A and Da Silva, Altigran S and Teixeira, Juliana S},
  journal={ACM SIGMOD Record},
  volume={31},
  number={2},
  pages={84--93},
  year={2002},
}

@article{laender2002debye,
  title={{DEByE}--Data Extraction by Example},
  author={Laender, Alberto HF and Ribeiro-Neto, Berthier and Da Silva, Altigran S},
  journal={Data \& Knowledge Engineering},
  volume={40},
  number={2},
  pages={121--154},
  year={2002},
  publisher={Elsevier}
}

@inproceedings{arifanto2018domain,
  title={Domain Specific Language for Web Scraper Development},
  author={Arifanto, Randy and Asnar, Yudistira DW and Liem, MM Inggriani},
  booktitle={2018 5th International Conference on Data and Software Engineering (ICoDSE)},
  pages={1--6},
  year={2018},
}

@Misc{pandas,
year={2024}, 
  howpublished = {\url{https://pandas.pydata.org}},
  note         = {Accessed: August 20, 2025},
  title        = {Pandas, a {Python} data analysis and manipulation tool.},
}

@inproceedings{foofah,
author = {Jin, Zhongjun and Anderson, Michael R. and Cafarella, Michael and Jagadish, H. V.},
title = {Foofah: Transforming Data By Example},
year = {2017},
doi = {10.1145/3035918.3064034},
publisher = {ACM},
booktitle = {Proceedings of the ACM SIGMOD International Conference on Management of Data},
pages = {683–698}
}

@ARTICLE{rigel,
  author={Chen, Ran and Weng, Di and Huang, Yanwei and Shu, Xinhuan and Zhou, Jiayi and Sun, Guodao and Wu, Yingcai},
  journal={IEEE Transactions on Visualization and Computer Graphics}, 
  title={Rigel: Transforming Tabular Data by Declarative Mapping}, 
  year={2023},
  volume={29},
  number={1},
  pages={128-138},
  doi={10.1109/TVCG.2022.3209385}}

@inproceedings{hesenius2025draw,
  title={How To Draw Commands? An Elicitation Study for Sketching on Spreadsheets},
  author={Hesenius, Marc and Krvavac, Mak and Valbj{\"o}rnsson, Valbj{\"o}rn J{\'o}n and Erika, Theresia Mita and Book, Matthias},
  booktitle={Proceedings of the CHI Conference on Human Factors in Computing Systems},
  pages={1--31},
  year={2025},
  doi = {10.1145/3706598.3715269},
}

@Misc{D3-js,
year={2024}, 
  howpublished = {\url{https://d3js.org/}},
  note         = {Accessed: August 20, 2025},
  title        = {{D3.js}, a web-standard JavaScript library for creating visualizations.},
}

@article{saket2016visualization,
  title={Visualization by Demonstration: An Interaction Paradigm for Visual Data Exploration},
  author={Saket, Bahador and Kim, Hannah and Brown, Eli T and Endert, Alex},
  journal={IEEE Transactions on Visualization and Computer Graphics},
  volume={23},
  number={1},
  pages={331--340},
  year={2016},
  publisher={IEEE},
  doi={10.1109/TVCG.2016.2598839}
}

@article{shi2022supporting,
  title={Supporting Expressive and Faithful Pictorial Visualization Design with Visual Style Transfer},
  author={Shi, Yang and Liu, Pei and Chen, Siji and Sun, Mengdi and Cao, Nan},
  journal={IEEE Transactions on Visualization and Computer Graphics},
  volume={29},
  number={1},
  pages={236--246},
  year={2022},
  publisher={IEEE},
  doi={10.1109/TVCG.2022.3209486}
}

@inproceedings{teng2021sketch2vis,
  title={Sketch2Vis: Generating Data Visualizations from Hand-drawn Sketches with Deep Learning},
  author={Teng, Zhongwei and Fu, Quchen and White, Jules and Schmidt, Douglas C},
  booktitle={IEEE International Conference on Machine Learning and Applications (ICMLA)},
  pages={853--858},
  year={2021},
  organization={IEEE}
}

@book{cypher1993watch,
  title={Watch What I Do: Programming by Demonstration},
  author={Cypher, Allen and Halbert, Daniel Conrad},
  year={1993},
  publisher={MIT Press}
}

@article{narechania2025provenancelens,
    title = {{Utilizing Provenance as an Attribute for Visual Data Analysis: A Design Probe with ProvenanceLens}},
    shorttitle = {{ProvenanceLens}},
    author = {{Narechania}, Arpit and {Guo}, Shunan and {Koh}, Eunyee and {Endert}, Alex and {Hoffswell}, Jane},
    journal = {IEEE Transactions on Visualization and Computer Graphics},
    year = {2025},
    doi = {10.1109/TVCG.2025.3571708},
    url = {https://doi.org/10.1109/TVCG.2025.3571708},
    publisher = {IEEE}
}

@book{halbert1984programming,
  title={Programming By Example},
  author={Halbert, Daniel Conrad},
  year={1984},
  publisher={University of California, Berkeley}
}

@article{yang2023fine,
  title={Fine-Grained Visual Prompting},
  author={Yang, Lingfeng and Wang, Yueze and Li, Xiang and Wang, Xinlong and Yang, Jian},
  journal={NeurIPS},
  volume={36},
  pages={24993--25006},
  year={2023}
}

@article{androutsopoulos2005natural,
  title={Natural Language Interaction},
  author={Androutsopoulos, Ion and Aretoulaki, Maria},
  year={2005}
}

@article{brooke1996sus,
  title={SUS-A quick and dirty usability scale},
  author={Brooke, John},
  journal={Usability Evaluation in Industry},
  volume={189},
  number={194},
  pages={4--7},
  year={1996},
  publisher={London, England}
}

@article{sperrle2024wizard,
  title={A Wizard of Oz Study of Guidance Strategies and Dynamics},
  author={Sperrle, Fabian and El-Assady, Mennatallah and Arleo, Alessio and Ceneda, Davide},
  journal={IEEE Transactions on Visualization and Computer Graphics},
  year={2024},
  publisher={IEEE},
  doi={10.1109/TVCG.2024.3418782}
}

@article{croisdale2025deckflow,
  title={{DeckFlow}: Iterative Specification on a Multimodal Generative Canvas},
  author={Croisdale, Gregory and Huang, Emily and Chung, John Joon Young and Guo, Anhong and Wang, Xu and Henley, Austin Z and Omar, Cyrus},
  journal={arXiv preprint arXiv:2506.15873},
  year={2025}
}

@inproceedings{deline2010code,
  title={Code canvas: zooming towards better development environments},
  author={DeLine, Robert and Rowan, Kael},
  booktitle={Proceedings of the 32nd ACM/IEEE International Conference on Software Engineering-Volume 2},
  pages={207--210},
  year={2010}
}

@article{yunita2022everything,
  title={‘Everything is data’: towards one big data ecosystem using multiple sources of data on higher education in Indonesia},
  author={Yunita, Ariana and Santoso, Harry B and Hasibuan, Zainal A},
  journal={Journal of Big Data},
  volume={9},
  number={1},
  pages={91},
  year={2022},
  publisher={Springer}
}

@inproceedings{huang2024table,
  title={Table Illustrator: Puzzle-Based Interactive Authoring of Plain Tables},
  author={Huang, Yanwei and Yang, Yurun and Shu, Xinhuan and Chen, Ran and Weng, Di and Wu, Yingcai},
  booktitle={Proceedings of the CHI Conference on Human Factors in Computing Systems},
  pages={1--18},
  year={2024},
  doi = {10.1145/3613904.3642415}
}

@inproceedings{pr-bias,
author = {Dell, Nicola and Vaidyanathan, Vidya and Medhi, Indrani and Cutrell, Edward and Thies, William},
title = {"Yours is better!": participant response bias in HCI},
year = {2012},
url = {https://doi.org/10.1145/2207676.2208589},
doi = {10.1145/2207676.2208589},
booktitle = {Proceedings of the CHI Conference on Human Factors in Computing Systems},
pages = {1321–1330},
}

@article{guidance-typology,
author = {Pérez-Messina, I. and Ceneda, D. and El-Assady, M. and Miksch, S. and Sperrle, F.},
title = {A Typology of Guidance Tasks in Mixed-Initiative Visual Analytics Environments},
journal = {Computer Graphics Forum},
volume = {41},
number = {3},
pages = {465-476},
doi = {https://doi.org/10.1111/cgf.14555},
url = {https://onlinelibrary.wiley.com/doi/abs/10.1111/cgf.14555},
eprint = {https://onlinelibrary.wiley.com/doi/pdf/10.1111/cgf.14555},
year = {2022}
}

@inproceedings{narechania2023datacockpit,
  title={{DataCockpit: A Toolkit for Data Lake Navigation and Monitoring Utilizing Quality and Usage Information}},
  author={Narechania, Arpit and Chakraborty, Surya and Agarwal, Shivam and Sinha, Atanu R and Rossi, Ryan A and Du, Fan and Hoffswell, Jane and Guo, Shunan and Koh, Eunyee and Endert, Alex and others},
  booktitle={2023 IEEE International Conference on Big Data (BigData)},
  pages={5305--5310},
  year={2023},
  organization={IEEE}
}

@inproceedings{shen2025prompting,
  title={Prompting generative AI with interaction-augmented instructions},
  author={Shen, Leixian and Li, Haotian and Wang, Yifang and Xie, Xing and Qu, Huamin},
  booktitle={Extended Abstracts of the CHI Conference on Human Factors in Computing Systems},
  pages={1--9},
  year={2025}
}

@inproceedings{xia2021ktabulator,
  title={{KTabulator}: Interactive ad hoc table creation using knowledge graphs},
  author={Xia, Siyuan and Anzum, Nafisa and Salihoglu, Semih and Zhao, Jian},
  booktitle={Proceedings of the CHI Conference on Human Factors in Computing Systems},
  pages={1--14},
  year={2021},
  doi = {10.1145/3411764.3445227}
}

@inproceedings{mayer2015user,
  title={User Interaction Models for Disambiguation in Programming by Example},
  author={Mayer, Mika{\"e}l and Soares, Gustavo and Grechkin, Maxim and Le, Vu and Marron, Mark and Polozov, Oleksandr and Singh, Rishabh and Zorn, Benjamin and Gulwani, Sumit},
  booktitle={Proceedings of the 28th Annual ACM Symposium on User Interface Software and Technology},
  pages={291--301},
  year={2015},
  doi = {10.1145/2807442.2807459},
}

@inproceedings{chopra2025challenges,
  title={Challenges in Using Conversational AI for Data Science},
  author={Chopra, Bhavya and Singha, Ananya and Fariha, Anna and Gulwani, Sumit and Parnin, Chris and Tiwari, Ashish and Henley, Austin Z},
  booktitle={Proceedings of the Workshop on Human-In-the-Loop Data Analytics},
  pages={1--7},
  year={2025}
}

@inproceedings{guo2024investigating,
author = {Guo, Jiajing and Mohanty, Vikram and Piazentin Ono, Jorge H and Hao, Hongtao and Gou, Liang and Ren, Liu},
title = {Investigating Interaction Modes and User Agency in Human-LLM Collaboration for Domain-Specific Data Analysis},
year = {2024},
url = {https://doi.org/10.1145/3613905.3651042},
doi = {10.1145/3613905.3651042},
booktitle = {Extended Abstracts of the CHI Conference on Human Factors in Computing Systems},
}

@inproceedings{narechania2021diy,
  title={DIY: Assessing the Correctness of Natural Language to SQL Systems},
  author={Narechania, Arpit and Fourney, Adam and Lee, Bongshin and Ramos, Gonzalo},
  booktitle={Proceedings of the 26th International Conference on Intelligent User Interfaces},
  pages={597--607},
  year={2021}
}

@article{yu2018spider,
  title={Spider: A large-scale human-labeled dataset for complex and cross-domain semantic parsing and text-to-sql task},
  author={Yu, Tao and Zhang, Rui and Yang, Kai and Yasunaga, Michihiro and Wang, Dongxu and Li, Zifan and Ma, James and Li, Irene and Yao, Qingning and Roman, Shanelle and others},
  journal={Proceedings of the  Conference on Empirical Methods in Natural Language Processing},
  year={2018}
}

@inproceedings{chopra2023cowrangler,
  title={{CoWrangler}: Recommender System for Data-Wrangling Scripts},
  author={Chopra, Bhavya and Fariha, Anna and Gulwani, Sumit and Henley, Austin Z and Perelman, Daniel and Raza, Mohammad and Shi, Sherry and Simmons, Danny and Tiwari, Ashish},
  booktitle={Companion of the 2023 International Conference on Management of Data},
  pages={147--150},
  year={2023}
}

@article{l2024learnable,
  title={Learnable and expressive visualization authoring through blended interfaces},
  author={L’Yi, Sehi and van den Brandt, Astrid and Adams, Etowah and Nguyen, Huyen N and Gehlenborg, Nils},
  journal={IEEE Transactions on Visualization and Computer Graphics},
  volume={31},
  number={1},
  pages={459--469},
  year={2024},
  publisher={IEEE},
  doi={10.1109/TVCG.2024.3456598}
}

@inproceedings{epperson2022leveraging,
  title={Leveraging Analysis History for
Improved In Situ Visualization Recommendation},
  author={Epperson, Will and Jung-Lin Lee, Doris and Wang, Leijie and Agarwal, Kunal and Parameswaran, Aditya G and Moritz, Dominik and Perer, Adam},
  booktitle={Computer Graphics Forum},
  volume={41},
  number={3},
  pages={145--155},
  year={2022},
  organization={Wiley Online Library}
}










\end{document}